\documentclass[superscriptaddress, reprint, amssymb, amsmath, aps, prl]{revtex4-1}
\usepackage{graphicx}
\usepackage{color}

\newcommand{\W}{\mathbf{W}} 
\newcommand{\Q}{\mathbf{Q}}
\newcommand{\beq}{\begin{equation}}
\newcommand{\eeq}{\end{equation}}
\newcommand{\arr}{\rightarrow}
\newcommand{\me}[3]{\langle #1 | #2 | #3 \rangle}

\newcommand{\ket}[1]{| #1 \rangle}
\newcommand{\average}[1]{\langle #1 \rangle}

\newcommand{\s}{\sigma}
\newcommand{\Z}{\mathcal{Z}}
\newcommand{\TP}{\text{TP}}
\newcommand{\RP}{\text{RP}}
\newcommand{\AP}{\text{AP}}

\begin{document}

\title{A path-based approach to random walks on networks \\ characterizes how proteins evolve new function}

\author{Michael Manhart}
\affiliation{Department of Physics and Astronomy, Rutgers University, Piscataway, NJ 08854}
\author{Alexandre V. Morozov}
 \email{morozov@physics.rutgers.edu}
\affiliation{Department of Physics and Astronomy, Rutgers University, Piscataway, NJ 08854}
\affiliation{BioMaPS Institute for Quantitative Biology, Rutgers University, Piscataway, NJ 08854}
\date{\today}

\begin{abstract}
We develop a path-based approach to continuous-time random walks on networks with arbitrarily weighted edges. We describe an efficient numerical algorithm for calculating statistical properties of the stochastic path ensemble. After demonstrating our approach on two reaction rate problems, we present a biophysical model that describes how proteins evolve new functions while maintaining thermodynamic stability. We use our methodology to characterize dynamics of evolutionary adaptation, reproducing several key features observed in directed evolution experiments. We find that proteins generally fall into two qualitatively different regimes of adaptation depending on their binding and folding energetics.
\end{abstract}

\maketitle



     Random walks on networks are ubiquitous across physics, chemistry, and biology, including molecular evolution \cite{Weinreich2006, Poelwijk2007, Carneiro2010}, protein folding \cite{Noe2009}, chemical reactions \cite{Bolhuis2002}, transport and search in complex media \cite{benAvraham2000, Condamin2007}, stochastic phenotypes \cite{Roma2005}, and cell-type differentiation \cite{Waddington1957, Kauffman1993, Enver2009}. Each node on the network is assigned a value of the objective function (for example, energy or fitness) which defines the rates of jumping to the neighboring nodes. Statistical properties of random walks determine quantities of interest such as mean first-passage times and path length distributions. Characterizing the diversity of stochastic paths is a central issue in evolutionary theory \cite{Weinreich2006, Poelwijk2007, Carneiro2010, Bridgham2009, Lobkovsky2011}.

     Analytical treatments of random walks on networks tend to be limited to simple models with equally weighted edges \cite{benAvraham2000, Noh2004, Bollt2005, Condamin2007}, while direct simulations can be computationally expensive, especially when rare events are considered.
In reaction rate theory, ensembles of stochastic trajectories may be built by transition path sampling \cite{Dellago1998, Dellago2003, Hummer2004, Harland2007}; however, this method  involves considerable computational costs in complex systems.  
Another alternative, transition path theory \cite{E2006, Metzner2009, Noe2009}, requires a numerical solution of the backward equation.  Neither approach directly addresses the diversity of stochastic paths.

     Here we develop a systematic and numerically efficient path-based approach to stochastic processes. Our method is applicable to semi-Markov jump processes (i.e., continuous-time random walks \cite{Weiss1994}) on networks with arbitrary edge weights.
The approach is well-suited for obtaining statistics that describe the diversity of paths, such as the distribution of path lengths and path entropy. We use it to study adaptive dynamics of proteins evolving a new function while maintaining thermodynamic stability \cite{Zeldovich2007, Tokuriki2008, Bridgham2009, Tokuriki2009, Bloom2009}, a phenomenon of central interest in both natural and directed evolution (the latter aimed at engineering proteins with novel enzymatic activities \cite{Bloom2009, Whitehead2012}).



     A semi-Markov process on the state space $\mathcal{S}$ is defined by a set of jump probabilities, $\me{\s'}{\Q}{\s}$ for $\s \arr \s'$ ($\s,\s'\in\mathcal{S}$), and waiting time distributions $\psi_\s(t)$, where $\psi_\s(t)$ is the PDF of waiting exactly time $t$ in state $\s$ before making a jump \cite{Weiss1994}. We assume that $\psi_\s(t)$ has finite mean $w(\s)$ for all $\s \in \mathcal{S}$.  Note that $\mathcal{S}$ equipped with the jump matrix $\Q$ defines a network with directed, weighted edges.

     Define a path $\varphi$ as a sequence of states $\{\s_0, \s_1, \ldots, \s_\ell\}$.  The time-independent probability of the system taking the path $\varphi$ is $\Pi[\varphi] = \pi(\s_0) \prod_{i=0}^{\ell-1} \me{\s_{i+1}}{\Q}{\s_i}$, where $\pi(\s_0)$ is the initial distribution. Let $\Phi$ be an ensemble of paths; for example, all first-passage paths from a set of initial states $\mathcal{S}_i$ to a set of final states $\mathcal{S}_f$.  The partition function for this ensemble is $\Z_\Phi = \sum_{\varphi \in \Phi} \Pi[\varphi]$ (note that $\Z_\Phi$ equals the normalization of the initial distribution $\pi(\s_0)$ by probability conservation), and the entropy is $S_\Phi = - \Z^{-1}_\Phi \sum_{\varphi \in \Phi} \Pi[\varphi] \log (\Pi[\varphi]/\Z_\Phi)$.  Let $\mathcal{L}[\varphi]$ be the length (number of jumps) of path $\varphi$, and let $\mathcal{T}[\varphi] = \sum_{i=0}^{\ell-1} w(\s_i)$ be the average time of the path.  We also define $\mathcal{T}_{\s}[\varphi] = \sum_{i=0}^{\ell-1} \delta_{\s \s_i} w(\s_i)$, the average time the path spends in state $\s$, and the indicator functional $\mathcal{I}_{\s}[\varphi]$, which equals 1 if $\varphi$ contains $\s$ and 0 otherwise.
     
    The average time of paths in the ensemble is then given by $\bar{\tau}_\Phi = \average{\mathcal{T}}_\Phi = \Z_\Phi^{-1} \sum_{\varphi \in \Phi} \mathcal{T}[\varphi] \Pi[\varphi]$.  The average path length is $\bar{\ell}_\Phi = \average{\mathcal{L}}_\Phi$, and the path length distribution is given by $\rho_\Phi(\ell) = \Z_\Phi^{-1} \sum_{\varphi \in \Phi} \delta(\ell - \mathcal{L}[\varphi]) \Pi[\varphi]$ \cite{Harland2007}.  Let $\ell_\Phi^\text{sd}$ be the standard deviation of path lengths, $\average{\mathcal{I}_\s}_\Phi$ the spatial density of paths (the probability of paths in $\Phi$ hitting state $\s$), and $\average{\mathcal{T}_\s}_\Phi/\bar{\tau}_\Phi$ the fraction of time spent in state $\s$.
     
     Let $\ket{\pi}$ be the vector of initial state probabilities and $\ket{\s}$ be the vector with $1$ at position $\s$ and $0$ otherwise.  For each step $\ell$ and intermediate state $\s$, we can recursively calculate $P_\ell(\s) = \me{\s}{\Q^\ell}{\pi}$ and $T_\ell(\s)$, the total probability and average time of all paths that end at $\s$ in $\ell$ steps:
     
\begin{eqnarray}
P_{\ell}(\s') &=& \sum_{\text{nn } \s \text{ of } \s'} \me{\s'}{\Q}{\s} P_{\ell - 1} (\s), \\
T_{\ell}(\s') &=& \sum_{\text{nn } \s \text{ of } \s'} \me{\s'}{\Q}{\s} \left[ T_{\ell - 1} (\s) + w(\s) P_{\ell - 1} (\s) \right], \nonumber
\end{eqnarray}

\noindent where $P_{0} (\s) = \pi (\s)$, $T_{0} (\s) = 0$, and the sums run over all nearest neighbors (nn) $\s$ of $\s'$ ($\s' \in \mathcal{S}_f$ are treated as absorbing states).
Therefore $\Z_\Phi = \sum_{\ell=1}^\infty \sum_{\s \in \mathcal{S}_f} P_\ell(\s)$ and

\beq
\rho_\Phi(\ell) = \frac{1}{\Z_\Phi} \sum_{\s \in \mathcal{S}_f} P_\ell(\s), \quad \bar{\tau}_\Phi = \frac{1}{\Z_\Phi} \sum_{\ell=1}^\infty \sum_{\s \in \mathcal{S}_f} T_\ell(\s).
\eeq

\noindent Other ensemble averages, such as $S_\Phi$, $\average{\mathcal{I}_\s}_\Phi$, $\average{\mathcal{I}_\s \mathcal{I}_{\s'}}_\Phi$, and $\average{\mathcal{T}_\s}_\Phi$, can be calculated similarly. Furthermore, we can calculate mean path divergence that characterizes the spatial diversity of the paths in $\Phi$ \cite{Lobkovsky2011}:

\beq
\mathcal{D}_\Phi = \sum_{\ell = 0}^\infty \sum_{\s,\s' \in \mathcal{S}} d(\s,\s') P_\ell(\s) P_\ell(\s'),
\label{eq:mpd}
\eeq
where $d(\s, \s')$ is a distance metric on $\mathcal{S}$.

     Our algorithm allows for very general definitions of the path ensemble $\Phi$ without having to explicitly enumerate paths. For instance, $\Phi$ can include paths that begin and end at arbitrary sets of states, or are disallowed from passing through arbitrary sets of intermediate states. Restriction to first-passage paths is also straightforward.
     The time complexity of our algorithm is $\mathcal{O}(\gamma NL)$ ($\mathcal{O}(\gamma N^2L)$ for $\mathcal{D}_\Phi$), where $\gamma$ is the average number of nn, $N$ is the number of states visited by paths in $\Phi$, and $L \sim \bar{\ell}_\Phi$ is the cutoff path length. For simple random walks, $\bar{\ell}_\Phi \sim N^{d_w/d_f}$ for $d_w > d_f$ and $\bar{\ell}_\Phi \sim N$ for $d_w \leq d_f$, where $d_w$ is the dimension of the walk and $d_f$ is the fractal dimension of the space~\cite{Bollt2005, Condamin2007}. Therefore, the algorithm scales as $\mathcal{O}(\gamma N^{1 + d_w/d_f})$ for $d_w > d_f$ and $\mathcal{O}(\gamma N^2)$ for $d_w \leq d_f$, automatically accounting for the sparseness of network connections.

\begin{figure}
\begin{center}
\begin{tabular}{ll}
\includegraphics[scale=0.21]{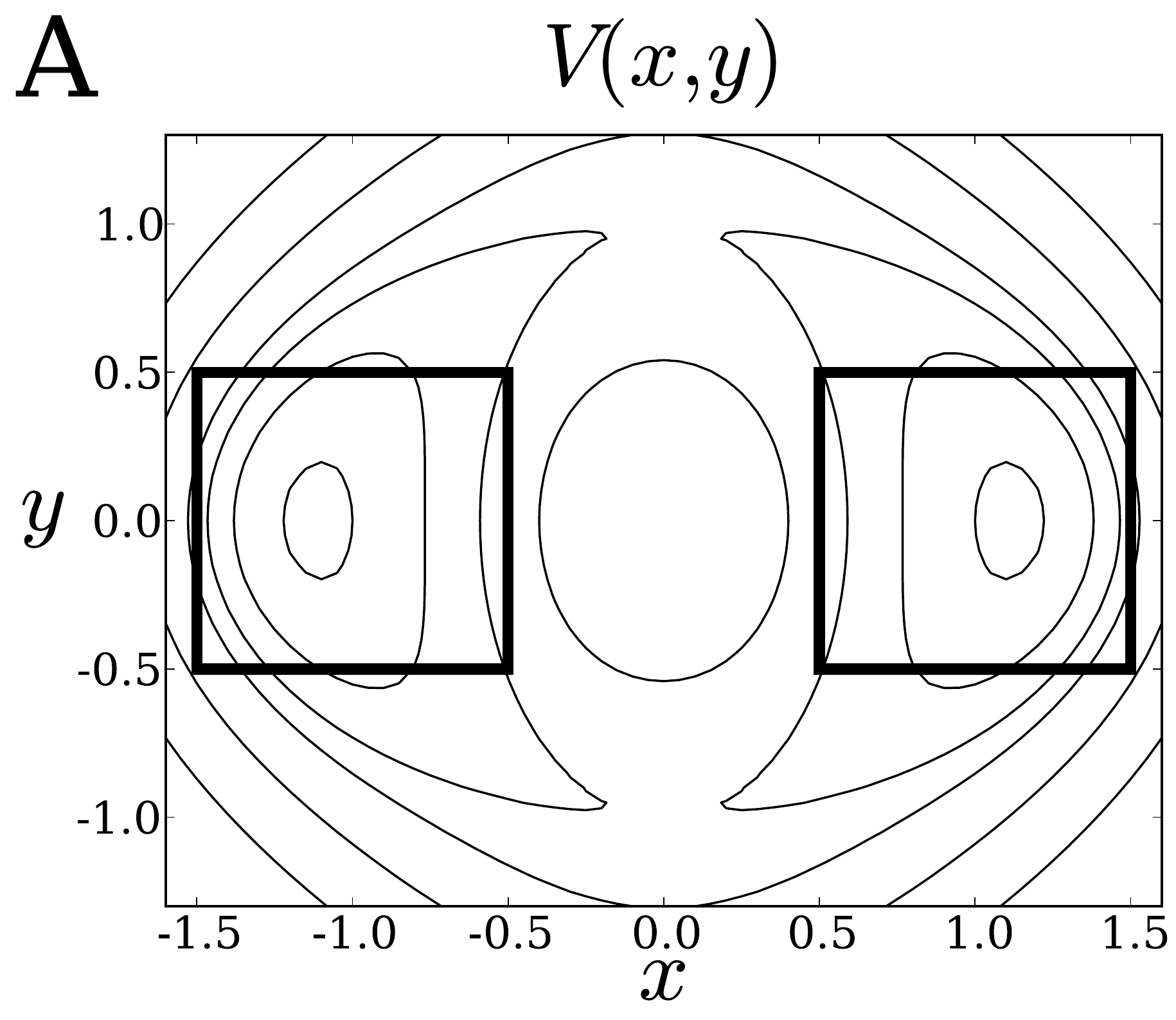} & \hspace{-0.3cm} \includegraphics[scale=0.24]{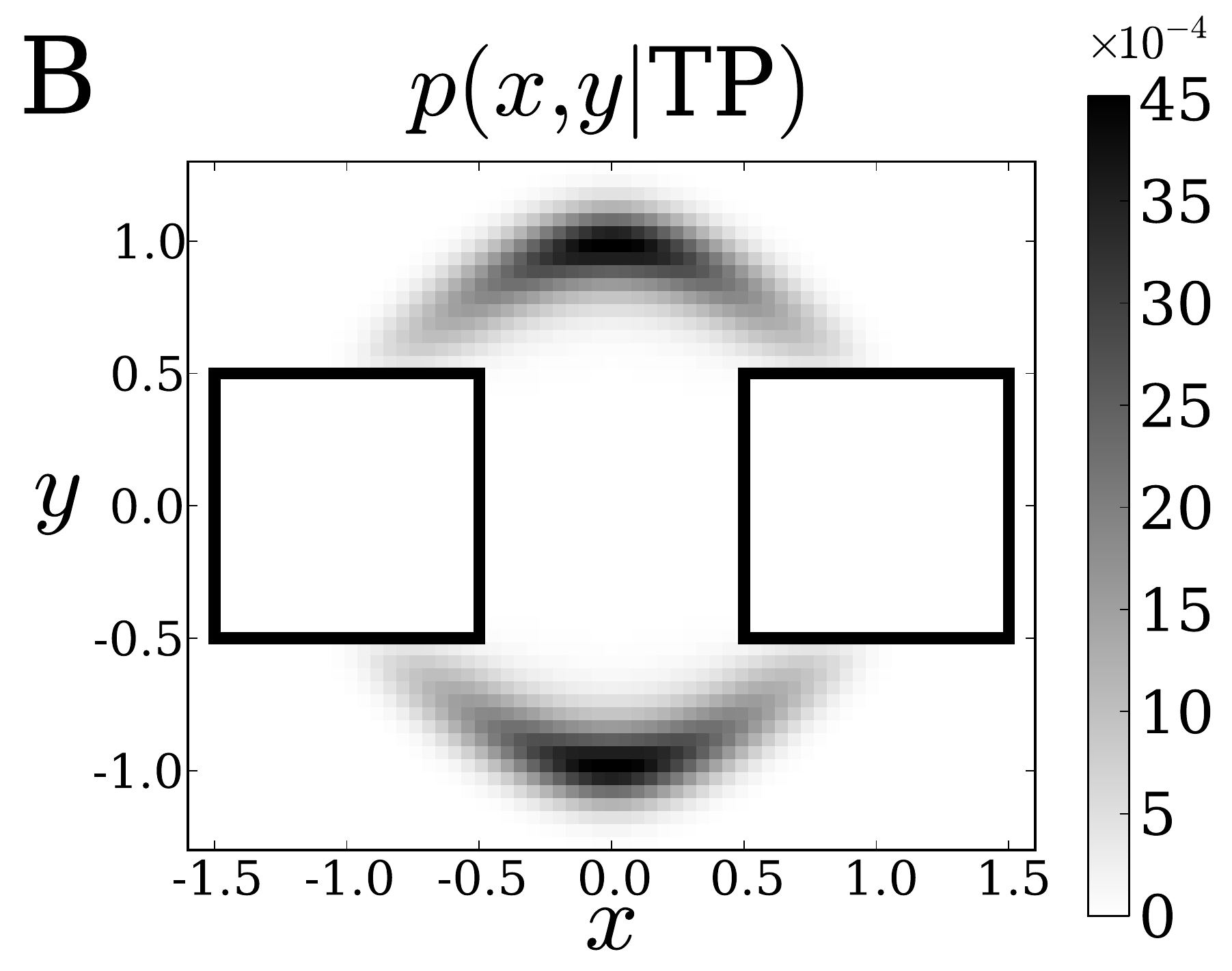} \\
\includegraphics[scale=0.2]{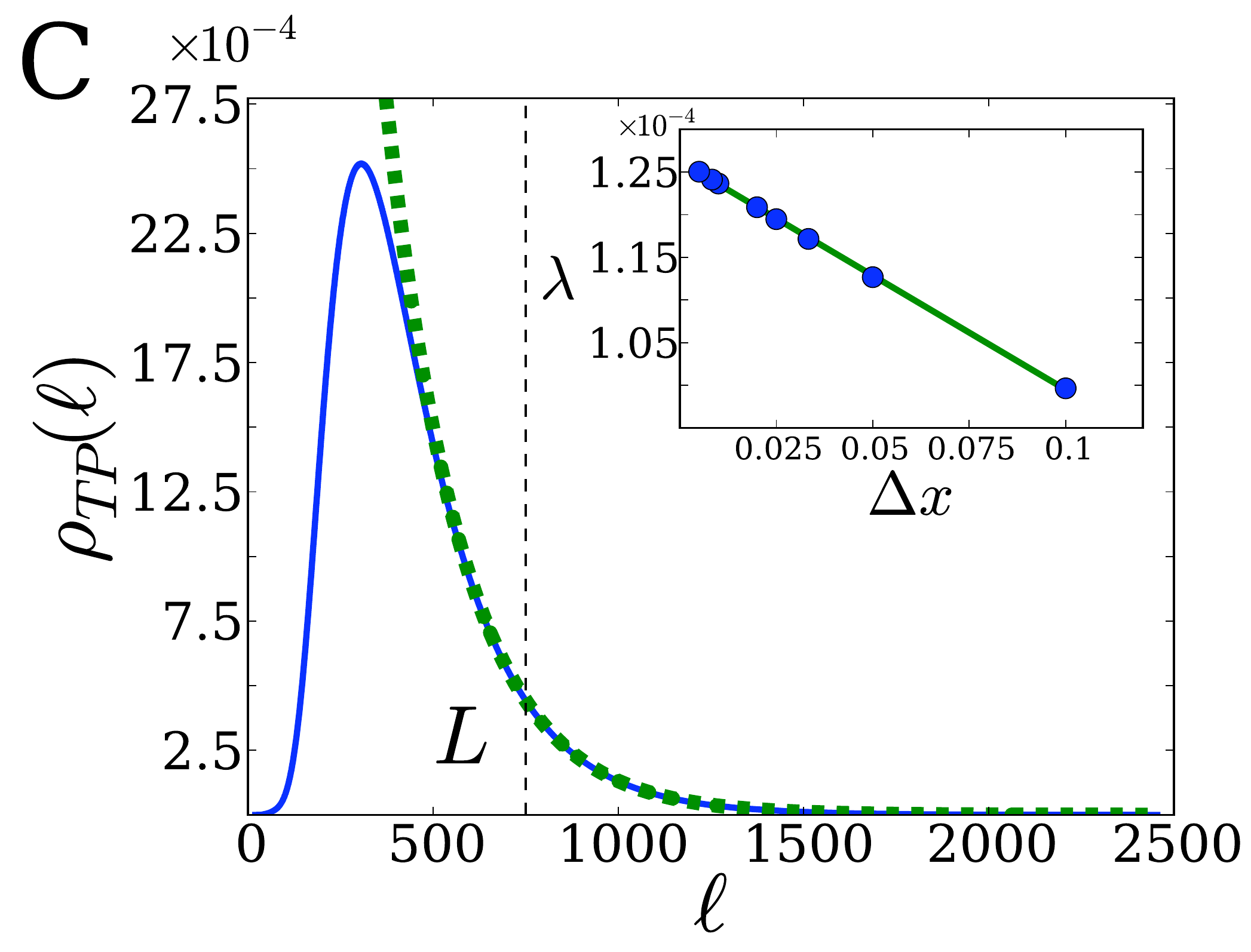} & \hspace{-0.4cm} \includegraphics[scale=0.2]{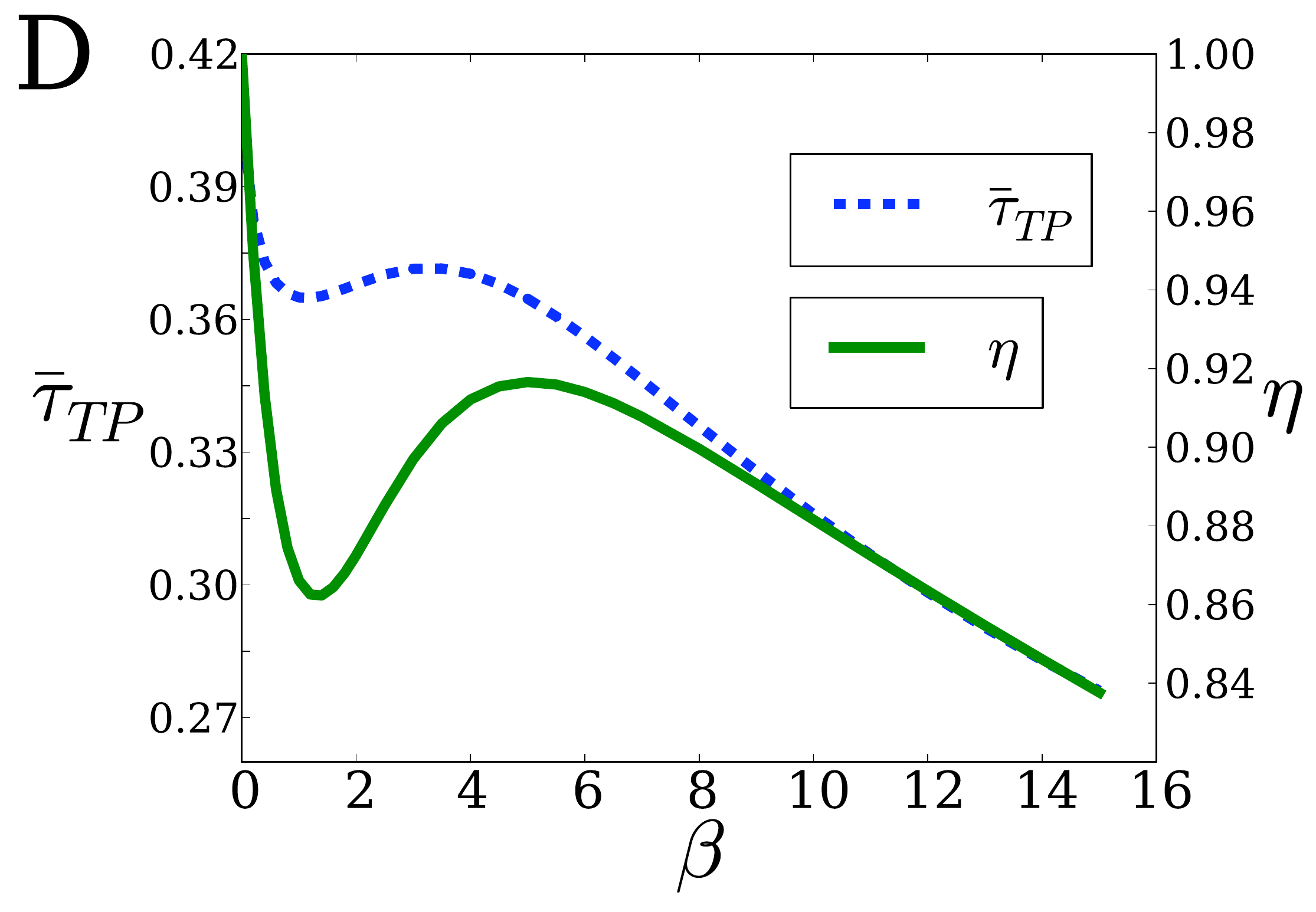} \\
\end{tabular}
\end{center}
\caption{(A) The double-well potential $V(x,y) = \frac{1}{6}(4(1 - x^2 - y^2)^2 + 2(x^2 - 2)^2 + ((x+y)^2 - 1)^2 + ((x-y)^2 - 1)^2 - 2)$ \cite{Dellago1998}. The space $\mathcal{S}$ is a square lattice on $[-1.6, 1.6] \times [-1.3, 1.3]$ with spacing $\Delta x$.
The metastable states are defined as $A = [-1.5,-0.5] \times [-0.5,0.5]$ and $B = [0.5,1.5] \times [-0.5,0.5]$.
(B) Density of states on TPs $p(x,y|\TP)$ for the double-well potential.
(C) Path length distribution $\rho_\TP(\ell)$ (solid, blue) and exponential fit in the interval $[L-50, L]$ (dashed, green), where $L = 750$.  In A,B,C, $\Delta x = 0.05$ and $\beta = 10$. Inset: TP flux $\lambda$ as a function of lattice spacing $\Delta x$.
(D) The relative mean path divergence $\eta = (\mathcal{D}_{\TP+\RP}(\beta)/\mathcal{D}_{\TP+\RP}(\beta = 0))^{1/2}$
and average time of TPs $\bar{\tau}_\TP$ versus $\beta$. The divergence $\eta$ is calculated using Eq.~\ref{eq:mpd} with $d(x,y;x',y') = (x-x')^2 + (y-y')^2$.
}
\label{fig:reaction_rates}
\end{figure}

     To determine the cutoff path length $L$, we recall that $\rho_\Phi(\ell) \sim e^{-\alpha \ell/\bar{\ell}_\Phi}$ for sufficiently large $\ell$, where $\alpha = \mathcal{O} (1)$ (Fig. \ref{fig:reaction_rates}C) \cite{Bollt2005}. Other path statistics, such as the average time $\bar{\tau}_\Phi(\ell)$ of paths up to length $\ell$ (Fig. S1A), also show exponential asymptotic behavior. Therefore in practice one need only consider paths with $\ell < L$ and infer the contributions of all longer paths from an exponential fit to the tail, which considerably improves the efficiency of the algorithm. This procedure takes advantage of the fact that information about longer paths is already contained in the structure of shorter paths.

     We now illustrate our approach on two reaction rate problems \cite{Hanggi1990}.
Consider two metastable states $A$ and $B$ with boundaries $\partial A$ and $\partial B$.  Let TP denote the ensemble of transition paths between $A$ and $B$: these paths begin on either $\partial A$ or $\partial B$ and end on the opposite boundary without crossing any boundaries in between \cite{Dellago2003, Hummer2004}. Similarly, RP denotes the ensemble of paths which return to the boundary on which they started.
The initial states on $\partial A$ and $\partial B$ are weighted by the equilibrium distribution $\pi(\s_0) = e^{-\beta V(\s_0)}/Z$ for a potential $V(\s_0)$, inverse temperature $\beta = 1/T$, and state-space partition function $Z = \sum_{\s_0 \in \mathcal{S}} e^{-\beta V(\s_0)}$. By definition, the first step of all TP and RP paths is from $\partial A$ or $\partial B$ to a point outside of $A$ and $B$, and the waiting time on $\partial A$ or $\partial B$ is zero.
     
     Many TP statistics, such as the distribution of path lengths $\rho_\TP(\ell)$, average time $\bar{\tau}_\TP$, mean path divergence $\mathcal{D}_{\TP+\RP}$, and the density of states $p(\s|\TP) = \average{\mathcal{T}_\s}_\TP/\bar{\tau}_\TP$, can be calculated straightforwardly with our method (Figs. \ref{fig:reaction_rates}, S1, S2).
We approximate the overall flux of TPs as the probability of being on a TP divided by the average time of a TP \cite{Hummer2004}:

\beq
\lambda \approx \frac{p(\TP)}{\bar{\tau}_\TP} =
\frac{(1 - \pi_A - \pi_B) \Z_\TP}{\Z_\TP \bar{\tau}_\TP + \Z_\RP \bar{\tau}_\RP},
\label{eq:lambda}
\eeq
where $\Z_\TP$ and $\Z_\RP$ are partition functions for transition and return paths, and $\pi_A$ and $\pi_B$ are the equilibrium probabilities of $A$ and $B$. The reaction rates are given by $k_{A \arr B} = \lambda/(2 \pi_A)$ and $k_{B \arr A} = \lambda/(2 \pi_B)$.


     First we consider a 2D double-well potential (Fig.~\ref{fig:reaction_rates}A), where $\mathcal{S}$ is a square lattice with spacing $\Delta x$ and jumps between nn have Monte Carlo rates $\me{x',y'}{\W}{x,y} = (\Delta x)^{-2} \min[1,e^{-\beta (V(x',y') - V(x,y))}]$.
Mean waiting times are given by $w(x,y) = (\sum_{\text{nn } (x',y') \text{ of } (x,y)} \me{x',y'}{\W}{x,y})^{-1}$, and jump probabilities are $\me{x',y'}{\Q}{x,y} = w(x,y) \me{x',y'}{\W}{x,y}$.
     
     The density of states on transition paths $p(x,y|\TP)$ shows two symmetric channels by which most reactions between $A$ and $B$ occur (Fig.~\ref{fig:reaction_rates}B).  As noted above, the distribution of path lengths $\rho_\TP(\ell)$ is exponential for long paths (Fig.~\ref{fig:reaction_rates}C).
In general, we expect paths to increase in length and diversity at higher temperatures. However, between $\beta = 5$ and $\beta = 1$ the paths become shorter and less diverse as $T$ increases (Figs.~\ref{fig:reaction_rates}D, S1B, S2). This is a signature of entropic switching~\cite{Metzner2006}: at a critical value of $\beta$, the two most energetically-favored pathways that dominated the low-$T$ behavior become less favorable than the shorter path through the middle. Entropic switching is reflected in plots of the relative path divergence, $\bar{\tau}_\TP$, $\bar{\ell}_\TP$, and $S_\TP$ (Figs.~\ref{fig:reaction_rates}D, S1B), which readily generalize to arbitrary network spaces.

     We can also calculate the continuous-space limit of the TP flux $\lambda$ (Eq.~\ref{eq:lambda}) and the reaction rates. We analytically continue $\lambda$ as a function of $\Delta x$: $\lambda(\Delta x) = \lambda_0 + \lambda_1 \Delta x + \mathcal{O}(\Delta x^2)$, where $\lambda_0$ is the continuous-limit flux and $\Delta x$ should be smaller then the smallest length scale of the potential.  Indeed, $\lambda (\Delta x)$ is linear (Fig.~\ref{fig:reaction_rates}C, inset), yielding continuous-limit rates of $k_{A \arr B} = k_{B \arr A} \approx 1.3 \times 10^{-4}$.  Therefore, one need only calculate $\lambda$ at a few finite lattice spacings to infer continuous-limit rates.  As shown in Fig. S3, our approach can be straightforwardly extended to reactions on more complex structures such as fractals, which serve as models of transport in disordered media~\cite{benAvraham2000}.

     We now apply our methodology to study evolution of protein function; here, the function is defined as binding a target such as an enzymatic substrate or another protein.  
Let $E_f$ be the protein folding free energy (i.e., the free energy difference between its folded and unfolded states), and $E_b$ the free energy of binding relative to the chemical potential of the target molecule. Then the protein has the probability of folding $1/(1 + e^{\beta E_f})$ and, independently, the probability of binding $1/(1 + e^{\beta E_b})$, where $\beta = 1.7$ (kcal/mol)$^{-1}$ is the inverse room temperature. We assume that the protein contributes fitness $1$ if it both folds and binds, and $f_0 < 1$ otherwise~\cite{Mayer2007}. Then fitness averaged over all proteins in an organism is given by
      
\beq
\mathcal{F}(E_f, E_b) = \frac{1 + f_0(e^{\beta E_f} + e^{\beta E_b} + e^{\beta(E_f + E_b)})}{(1 + e^{\beta E_f})(1 + e^{\beta E_b})}.
\label{eq:fitness}
\eeq

     The folding and binding energies are functions of the amino acid sequence $\s$. We assume that the protein has a small number $L$ of ``hotspot'' residues at the binding interface~\cite{Clackson1995}, and that each residue makes an additive contribution to the total energy~\cite{Serrano1993}: $E_f(\s) = E_f^0 + \sum_{i = 1}^L \epsilon_f(i, \s_i)$, $E_b(\s) = E_b^0 + \sum_{i = 1}^L \epsilon_b(i, \s_i)$, where $E_f^0,E_b^0$ are overall offsets and $\epsilon_{f,b}(i, \s_i)$ is the energy of amino acid $\s_i$ at position $i$.  The offset $E_f^0$ is a fixed contribution to folding energy from the rest of the protein, which we assume to be perfectly adapted; $\epsilon_f$'s are sampled from a Gaussian with mean 1.25 kcal/mol and standard deviation 1.6 kcal/mol \cite{Tokuriki2007}.  Since binding hotspots typically have a minimum penalty of 1-3 kcal/mol for mutations away from the wild-type amino acid~\cite{Bogan1998}, we set $\epsilon_b(i, \s_i^\text{best}) = 0 ~\forall ~i$ ($\s^\text{best}$ is the best-binding sequence: $E_b (\s^\text{best}) = E_b^0$), and sample the rest of $\epsilon_b$'s from an exponential distribution defined in the range of $(1,\infty)$ kcal/mol, with 2 kcal/mol mean~\cite{Thorn2001}. 
Here we consider $L=5$ hotspot residues and a reduced alphabet of 8 amino acids grouped by physico-chemical properties, resulting in $8^5 = 32768$ unique sequences.
The exact choices of these parameters have little effect on the overall qualitative features of the model.

     
     Our model naturally incorporates tradeoffs between function and stability~\cite{Tokuriki2009, Bloom2009, Wang2002}, even though binding and folding are uncorrelated~\cite{Tokuriki2008}.  Furthermore, our fitness landscape is nonlinear and thus epistatic: the fitness effect of a given mutation depends on the entire background sequence~\cite{Weinreich2006, Poelwijk2007, Carneiro2010}.  However, our landscape is correlated ($k^L$ sequences are determined by $2Lk$ $\epsilon_{f,b}$ parameters) and thus differs from completely random landscapes~\cite{Kauffman1993} in a manner consistent with experimental studies~\cite{Carneiro2010, Lobkovsky2011}.  
     
\begin{figure}
\begin{center}
\includegraphics[scale=0.21]{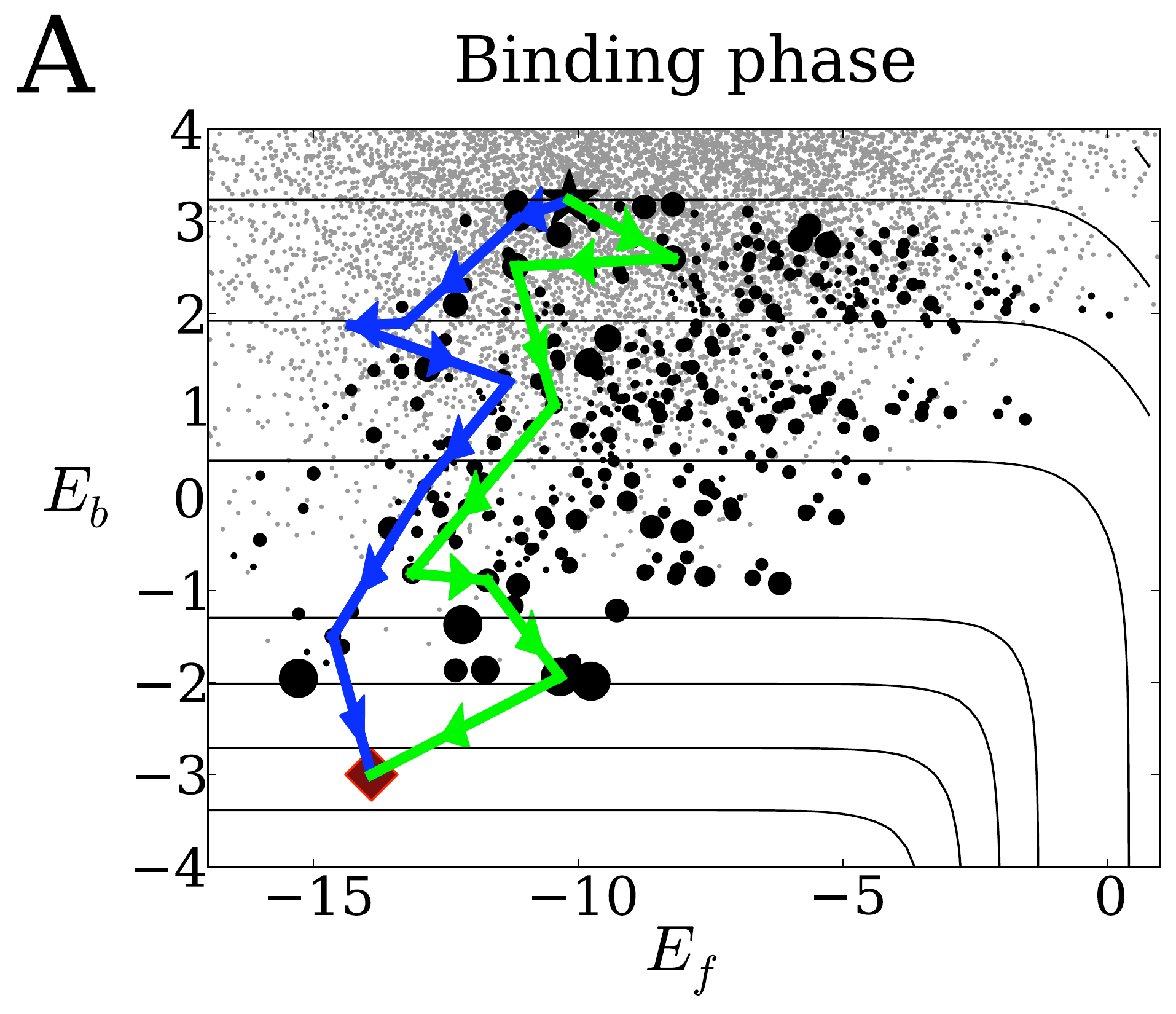} \hspace{-0.2cm}
\includegraphics[scale=0.21]{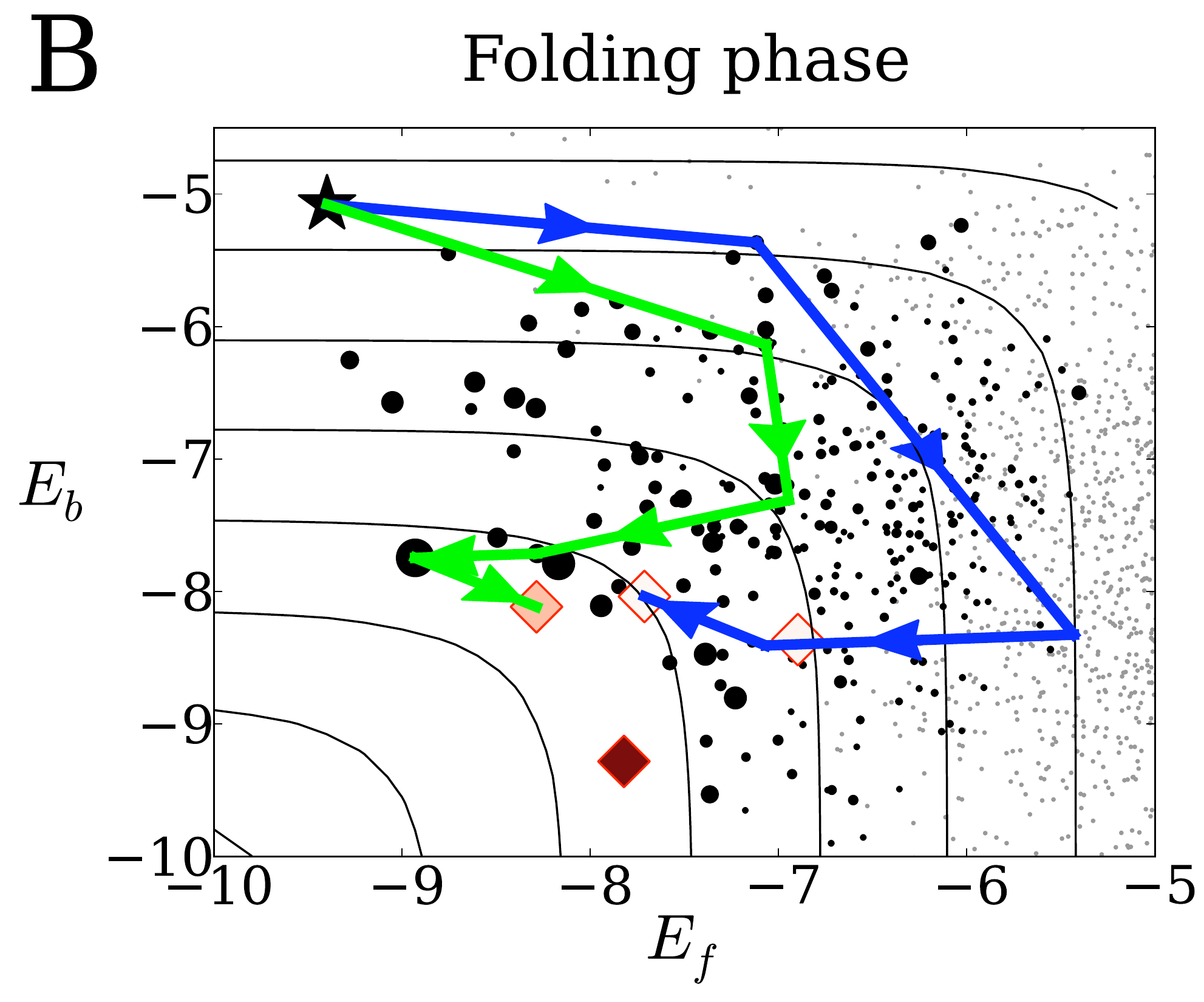}
\end{center}
\caption{Two realizations of the fitness landscape. (A) Binding phase, with $E_f^0 = -17$~kcal/mol and $E_b^0 = -3$~kcal/mol. (B)~Folding phase, with $E_f^0 = -3$~kcal/mol and $E_b^0 = -17$~kcal/mol ($E_f (\s^\text{best}) = -9.4$~kcal/mol).  Representative APs are shown in blue and green. Black star: sequence with global maximum on $\mathcal{F}_1$; red diamonds: local maxima on $\mathcal{F}_2$ shaded according to their commitment probabilities (i.e., probabilities to be reached from the initial state); black circles: intermediate states along APs, sized proportional to the density of APs $\average{\mathcal{I}_\s}_\AP$; small gray circles: states inaccessible to APs; black lines: contours of constant fitness.}
\label{fig:landscapes}
\end{figure}

     We sample one set of $\epsilon_f$'s and two sets of $\epsilon_{b}$'s for the old binding target and the new one. This procedure defines two fitness landscapes, $\mathcal{F}_1$ and $\mathcal{F}_2$ through Eq.~\ref{eq:fitness} ($E_f^0$ and $E_b^0$ are fixed). At first, each organism in the population has the sequence with maximum fitness under $\mathcal{F}_1$. The population then adapts to binding the new target on $\mathcal{F}_2$.  We assume the strong-selection limit: the population can only undergo substitutions that increase fitness.  This limit implies that the population is monomorphic: mutations arise one at a time and either disappear or fix rapidly \cite{Crow1970, Manhart2012}. Beneficial substitutions occur at a rate $Nu \ll 1$, where $N$ is the effective population size and $u$ is the mutation rate per amino acid.
Assuming Markovian waiting times, the jump probabilities are $\me{\s'}{\Q}{\s}=1/b(\s)$ if $\mathcal{F}(\s') > \mathcal{F}(\s)$ and 0 otherwise, where $b(\s)$ is the number of beneficial substitutions possible from $\s$.  Note that in this limit our results are independent of $f_0$ and $Nu$ only affects the overall time scale. The path ensemble consists of all adaptive paths (APs). Figure \ref{fig:landscapes} shows two realizations of $\mathcal{F}_2$ with representative APs.
            
\begin{figure}
\begin{center}
\begin{tabular}{ll}
\includegraphics[scale=0.21]{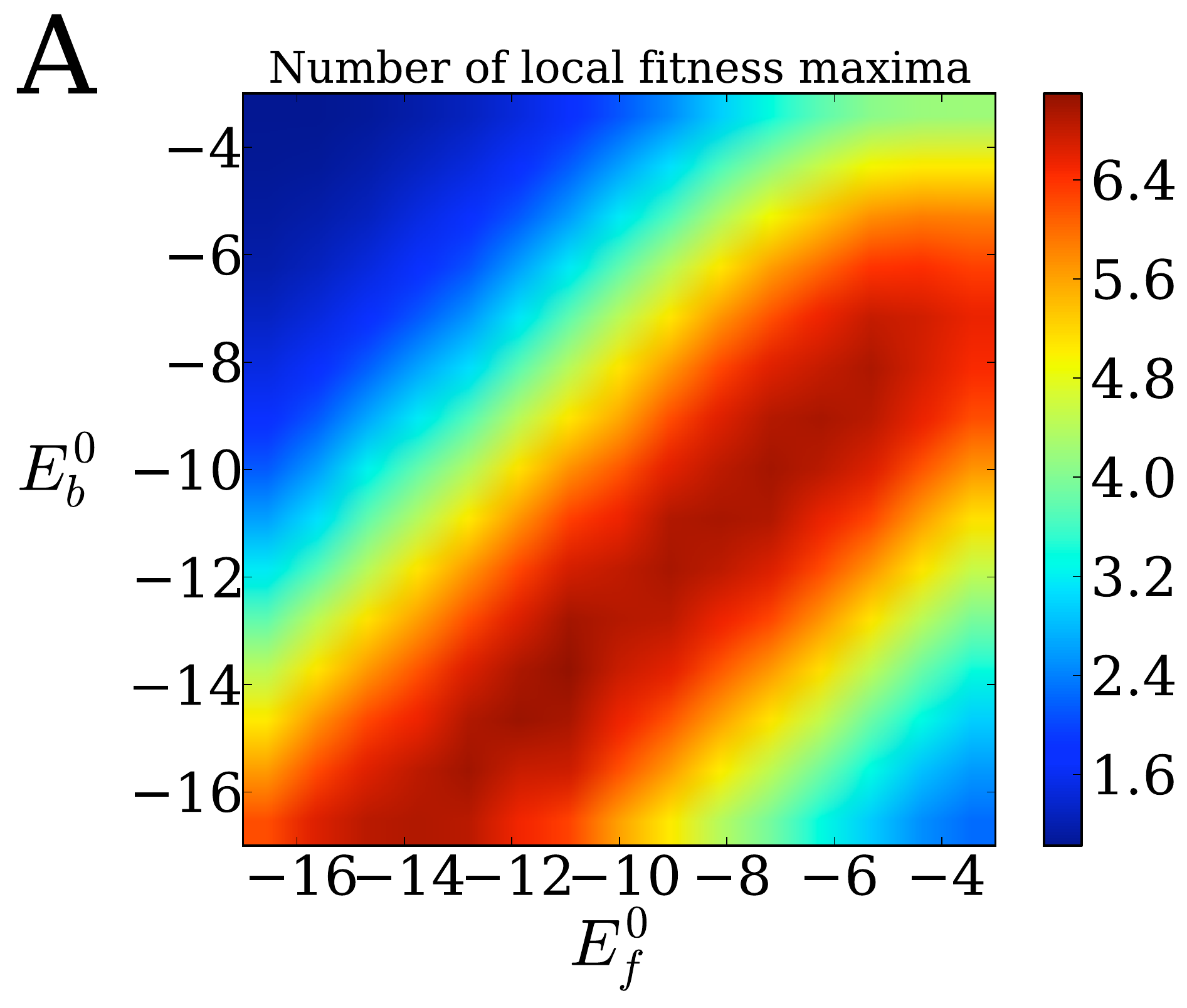} & \includegraphics[scale=0.2]{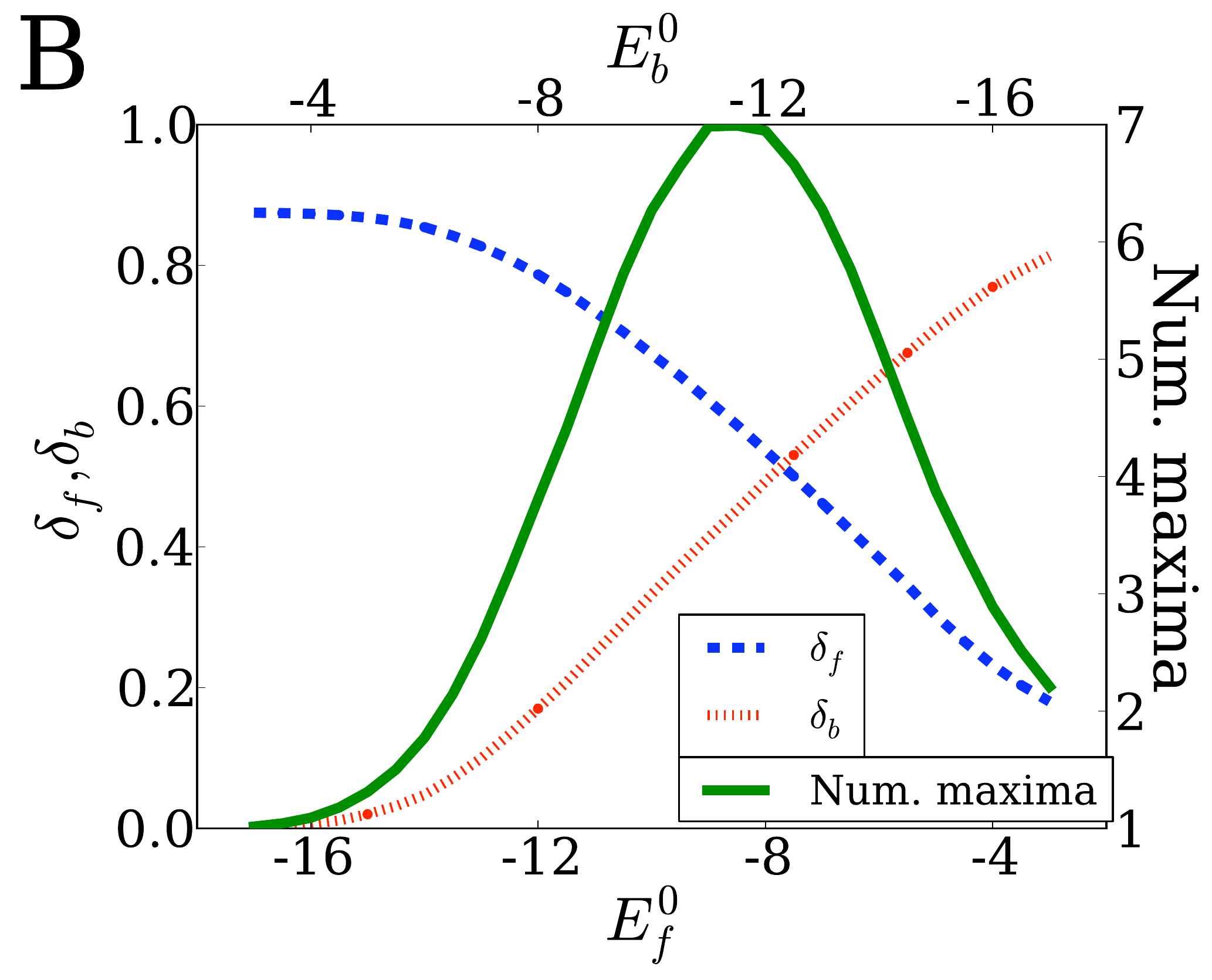} \\
\includegraphics[scale=0.2]{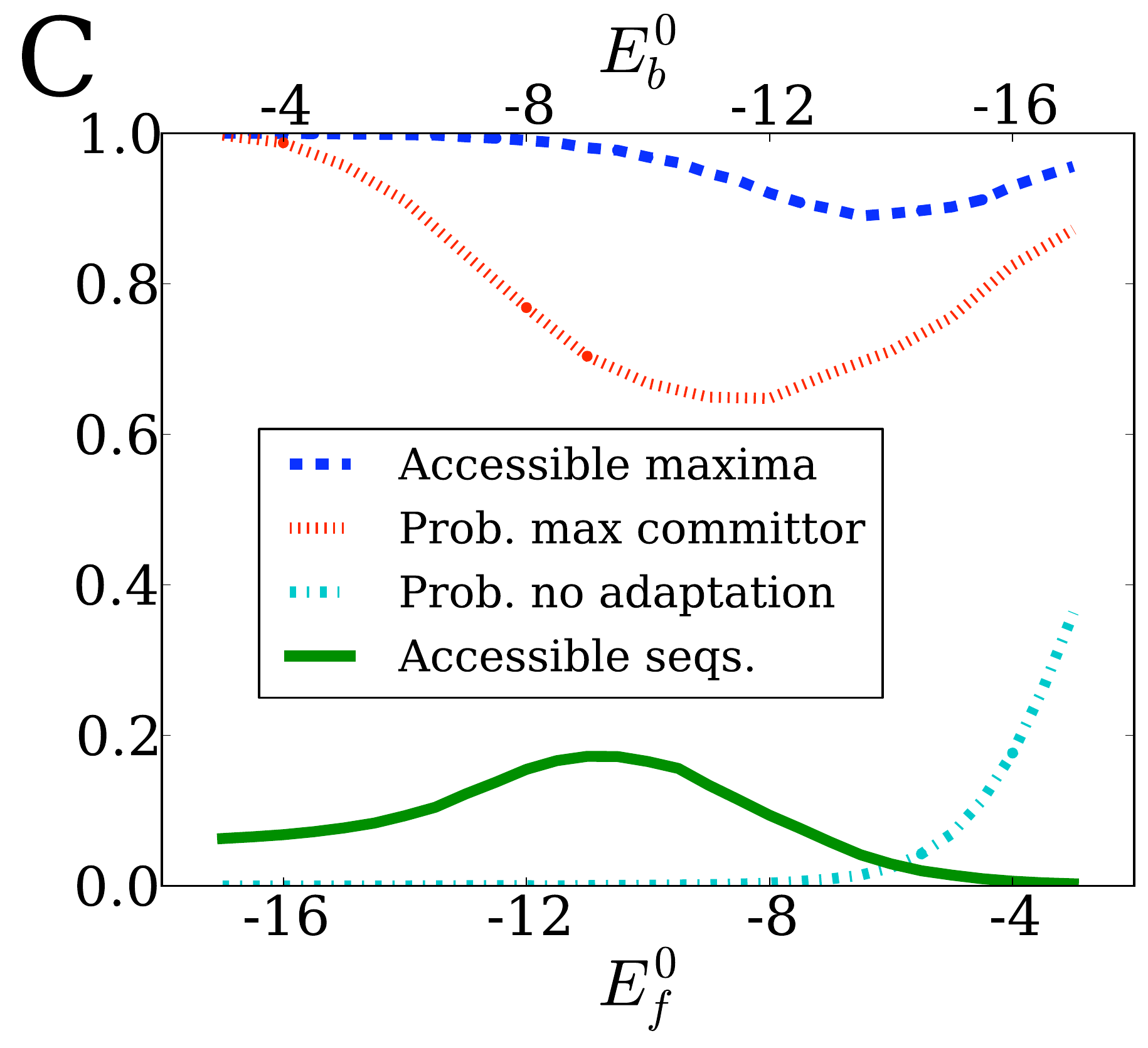} & \includegraphics[scale=0.2]{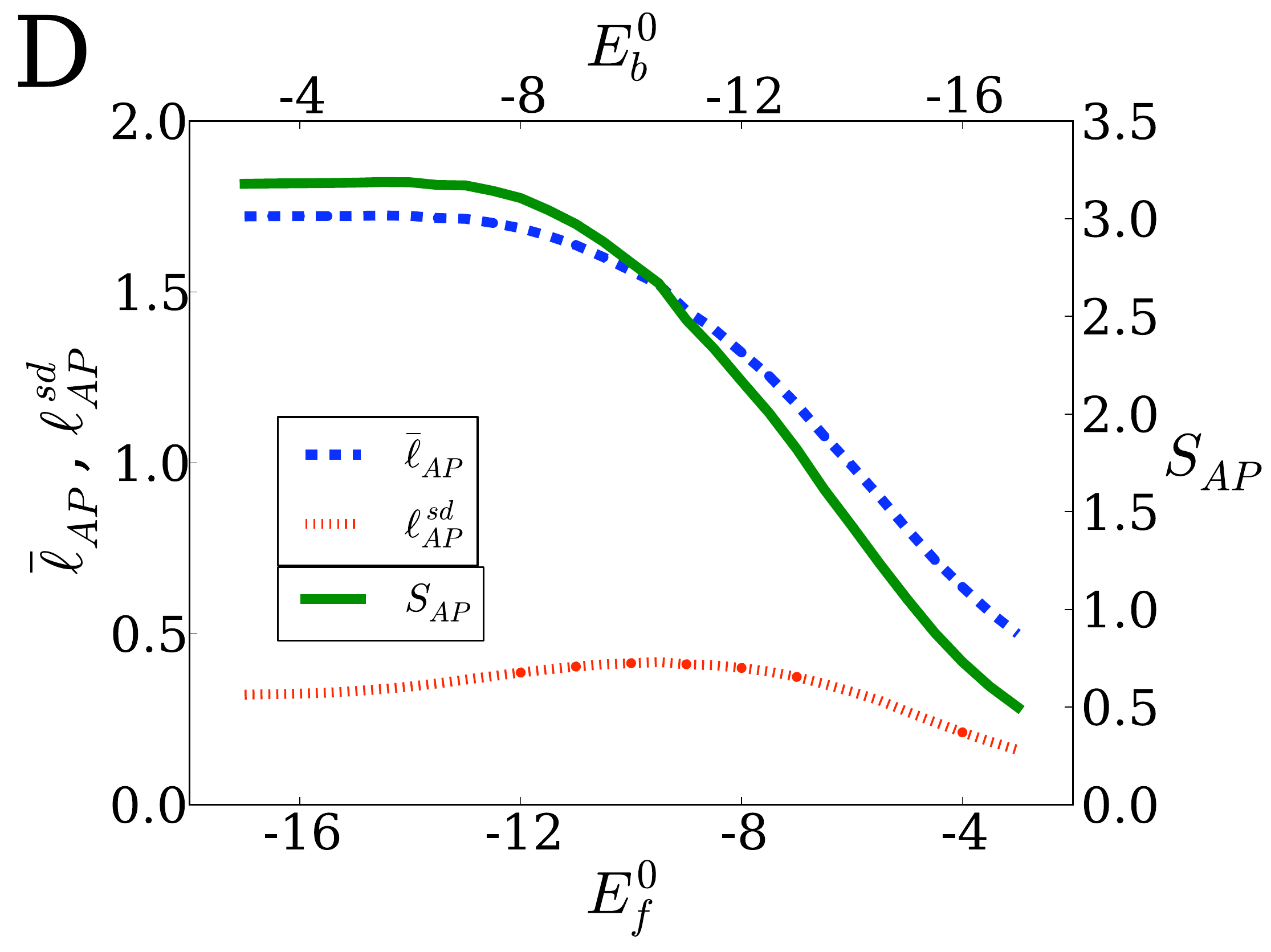} \\
\end{tabular}
\end{center}
\caption{
(A) Average number of local fitness maxima as a function of the energy offsets $E_f^0$ and $E_b^0$.
(B) Average number of local fitness maxima (solid, green), the average distance $\delta_f$ between the maxima and the best-folding sequence (dashed, blue), and the average distance $\delta_b$ between the maxima and the best-binding sequence (dotted, red) for the parameter subspace $E_f^0 + E_b^0 = -20$~kcal/mol. Note that the distance between two random sequences is $1 - 1/k$, where $k$ is the size of the alphabet.
(C) Fraction of local fitness maxima accessible from the initial state (dashed, blue), probability that the global maximum has the largest commitment probability among all local maxima (dotted, red), probability the initial sequence starts at a local maximum resulting in no adaptation (dashed and dotted, cyan), and fraction of state space accessible to APs (solid, green).
(D) Mean length $\bar{\ell}_\AP$, standard deviation $\ell_\AP^\text{sd}$, and entropy $S_\AP$. All quantities are per-residue.
The probability of no adaptation in (C) is an average over $2 \times 10^4$ landscape realizations; all other data points are averages over $5 \times 10^3$ realizations, and realizations with no adaptation are excluded.}
\label{fig:phases}
\end{figure}

     Since our fitness landscapes (Eq.~\ref{eq:fitness}) are randomly generated, we focus on their generic properties averaged over many realizations of $\epsilon_f$ and $\epsilon_{b}$ (Figs.~\ref{fig:phases}, S4). Our scans of the $E_f^0$-$E_b^0$ parameter space
reveal the existence of two qualitatively different phases of adaptation.  One, which we call the \emph{binding phase}, is observed when $E_f^0$ is low and $E_b^0$ is high (see Fig.~\ref{fig:landscapes}A for an example). In this case, the mean number of local fitness maxima is very low (Fig.~\ref{fig:phases}A,B) and $\delta_f$, the average Hamming distance between these maxima and the best-folding sequence (with the lowest $E_f$), is large (Fig.~\ref{fig:phases}B). In contrast, $\delta_b$, the average Hamming distance to the best-binding sequence, is close to zero. Thus in this phase the need to bind dominates adaptation.

     In the opposite limit (high $E_f^0$ and low $E_b^0$; see Fig.~\ref{fig:landscapes}B for an example), the \emph{folding phase} is observed in which the mean number of local maxima is also low (Fig.~\ref{fig:phases}A,B) but these maxima are much closer to the best-folding rather than the best-binding sequence (Fig.~\ref{fig:phases}B). Here, the need to preserve protein stability dominates adaptive dynamics. In the crossover regime between these two phases, there are many local maxima and therefore the most epistasis. This is reflected in the fact that the fraction of local maxima accessible from the initial state and the probability that the global maximum has the largest commitment probability are lower, while the fraction of sequence space accessible to APs is higher in this regime compared to the binding and folding phases (Fig.~\ref{fig:phases}C). In the crossover regime, the tradeoff between binding and folding alone can result in proteins with marginal folding stability, in contrast with previous hypotheses that explain marginal stability with mutational entropy \cite{Zeldovich2007} or a fitness function that disfavors hyperstable proteins \cite{DePristo2005}.
     
     On average, paths in the binding phase are longer than those in the folding phase, and adaptation takes more time (Figs.~\ref{fig:phases}D, S4A). Paths in this regime have higher entropy, indicating that adaptation involves a diverse set of intermediate sequences rather than a few dominant pathways. The standard deviation of path lengths is also higher (Fig.~\ref{fig:phases}D).    
In the folding phase APs tend to be short since the initial sequence is often either close to, or already at a local maximum (Figs.~\ref{fig:phases}C, S4B).
A similar situation is observed in directed evolution experiments where the initial sequence already has some affinity for the new ligand but cannot increase it any further \cite{Bershtein2008, Bridgham2009}.  In such cases, the sequences must first be mutated away from the local maximum. Furthermore, in the folding phase folding energy tends to increase at the beginning of paths and decrease toward the end, as a consequence of the distribution of sequences in energy space relative to the fitness contours (Fig.~\ref{fig:landscapes}B).  This is consistent with experiments in which folding stability is sacrificed first and recovered later en route to the new function~\cite{Bloom2009}.

     Our model can be extended to account for binding-mediated stability, in which binding stabilizes an otherwise disordered protein \cite{Brown2011}. We can also incorporate chaperone-assisted folding \cite{Rutherford2003} by modifying $E_f^0$ or the distribution of $\epsilon_f$'s. Furthermore, we can include ``folding hotspots'' away from the binding interface to see if they acquire stabilizing mutations as a buffer against destabilizing but function-improving mutations at the interface \cite{Tokuriki2009, Bloom2009}. The role of neutral and slightly deleterious mutations can be studied as well by using substitution rates from more complex population genetics models \cite{Crow1970, Manhart2012}, although we expect non-adaptive substitutions to play little role on short time scales. We look forward to studying these extensions in future work.


\begin{acknowledgments}
A.V.M. acknowledges support from National Institutes of Health (R01 HG004708) and an Alfred P. Sloan Research Fellowship.
\end{acknowledgments}

\bibliography{Submission1_text}

\begin{thebibliography}{42}%
\makeatletter
\providecommand \@ifxundefined [1]{%
 \@ifx{#1\undefined}
}%
\providecommand \@ifnum [1]{%
 \ifnum #1\expandafter \@firstoftwo
 \else \expandafter \@secondoftwo
 \fi
}%
\providecommand \@ifx [1]{%
 \ifx #1\expandafter \@firstoftwo
 \else \expandafter \@secondoftwo
 \fi
}%
\providecommand \natexlab [1]{#1}%
\providecommand \enquote  [1]{``#1''}%
\providecommand \bibnamefont  [1]{#1}%
\providecommand \bibfnamefont [1]{#1}%
\providecommand \citenamefont [1]{#1}%
\providecommand \href@noop [0]{\@secondoftwo}%
\providecommand \href [0]{\begingroup \@sanitize@url \@href}%
\providecommand \@href[1]{\@@startlink{#1}\@@href}%
\providecommand \@@href[1]{\endgroup#1\@@endlink}%
\providecommand \@sanitize@url [0]{\catcode `\\12\catcode `\$12\catcode
  `\&12\catcode `\#12\catcode `\^12\catcode `\_12\catcode `\%12\relax}%
\providecommand \@@startlink[1]{}%
\providecommand \@@endlink[0]{}%
\providecommand \url  [0]{\begingroup\@sanitize@url \@url }%
\providecommand \@url [1]{\endgroup\@href {#1}{\urlprefix }}%
\providecommand \urlprefix  [0]{URL }%
\providecommand \Eprint [0]{\href }%
\providecommand \doibase [0]{http://dx.doi.org/}%
\providecommand \selectlanguage [0]{\@gobble}%
\providecommand \bibinfo  [0]{\@secondoftwo}%
\providecommand \bibfield  [0]{\@secondoftwo}%
\providecommand \translation [1]{[#1]}%
\providecommand \BibitemOpen [0]{}%
\providecommand \bibitemStop [0]{}%
\providecommand \bibitemNoStop [0]{.\EOS\space}%
\providecommand \EOS [0]{\spacefactor3000\relax}%
\providecommand \BibitemShut  [1]{\csname bibitem#1\endcsname}%
\let\auto@bib@innerbib\@empty
\bibitem [{\citenamefont {Weinreich}\ \emph {et~al.}(2006)\citenamefont
  {Weinreich}, \citenamefont {Delaney}, \citenamefont {DePristo},\ and\
  \citenamefont {Hartl}}]{Weinreich2006}%
  \BibitemOpen
  \bibfield  {author} {\bibinfo {author} {\bibfnamefont {D.~M.}\ \bibnamefont
  {Weinreich}}, \bibinfo {author} {\bibfnamefont {N.~F.}\ \bibnamefont
  {Delaney}}, \bibinfo {author} {\bibfnamefont {M.~A.}\ \bibnamefont
  {DePristo}}, \ and\ \bibinfo {author} {\bibfnamefont {D.~L.}\ \bibnamefont
  {Hartl}},\ }\href@noop {} {\bibfield  {journal} {\bibinfo  {journal}
  {Science}\ }\textbf {\bibinfo {volume} {312}},\ \bibinfo {pages} {111}
  (\bibinfo {year} {2006})}\BibitemShut {NoStop}%
\bibitem [{\citenamefont {Poelwijk}\ \emph {et~al.}(2007)\citenamefont
  {Poelwijk}, \citenamefont {Kiviet}, \citenamefont {Weinreich},\ and\
  \citenamefont {Tans}}]{Poelwijk2007}%
  \BibitemOpen
  \bibfield  {author} {\bibinfo {author} {\bibfnamefont {F.~J.}\ \bibnamefont
  {Poelwijk}}, \bibinfo {author} {\bibfnamefont {D.~J.}\ \bibnamefont
  {Kiviet}}, \bibinfo {author} {\bibfnamefont {D.~M.}\ \bibnamefont
  {Weinreich}}, \ and\ \bibinfo {author} {\bibfnamefont {S.~J.}\ \bibnamefont
  {Tans}},\ }\href@noop {} {\bibfield  {journal} {\bibinfo  {journal} {Nature}\
  }\textbf {\bibinfo {volume} {445}},\ \bibinfo {pages} {383} (\bibinfo {year}
  {2007})}\BibitemShut {NoStop}%
\bibitem [{\citenamefont {Carneiro}\ and\ \citenamefont
  {Hartl}(2010)}]{Carneiro2010}%
  \BibitemOpen
  \bibfield  {author} {\bibinfo {author} {\bibfnamefont {M.}~\bibnamefont
  {Carneiro}}\ and\ \bibinfo {author} {\bibfnamefont {D.~L.}\ \bibnamefont
  {Hartl}},\ }\href@noop {} {\bibfield  {journal} {\bibinfo  {journal} {Proc.
  Natl. Acad. Sci. USA}\ }\textbf {\bibinfo {volume} {107}},\ \bibinfo {pages}
  {1747} (\bibinfo {year} {2010})}\BibitemShut {NoStop}%
\bibitem [{\citenamefont {No{\'{e}}}\ \emph {et~al.}(2009)\citenamefont
  {No{\'{e}}}, \citenamefont {Sch{\"{u}}tte}, \citenamefont {Vanden-Eijnden},
  \citenamefont {Reich},\ and\ \citenamefont {Weikl}}]{Noe2009}%
  \BibitemOpen
  \bibfield  {author} {\bibinfo {author} {\bibfnamefont {F.}~\bibnamefont
  {No{\'{e}}}}, \bibinfo {author} {\bibfnamefont {C.}~\bibnamefont
  {Sch{\"{u}}tte}}, \bibinfo {author} {\bibfnamefont {E.}~\bibnamefont
  {Vanden-Eijnden}}, \bibinfo {author} {\bibfnamefont {L.}~\bibnamefont
  {Reich}}, \ and\ \bibinfo {author} {\bibfnamefont {T.~R.}\ \bibnamefont
  {Weikl}},\ }\href@noop {} {\bibfield  {journal} {\bibinfo  {journal} {Proc.
  Natl. Acad. Sci. USA}\ }\textbf {\bibinfo {volume} {106}},\ \bibinfo {pages}
  {19011} (\bibinfo {year} {2009})}\BibitemShut {NoStop}%
\bibitem [{\citenamefont {Bolhuis}\ \emph {et~al.}(2002)\citenamefont
  {Bolhuis}, \citenamefont {Chandler}, \citenamefont {Dellago},\ and\
  \citenamefont {Geissler}}]{Bolhuis2002}%
  \BibitemOpen
  \bibfield  {author} {\bibinfo {author} {\bibfnamefont {P.~G.}\ \bibnamefont
  {Bolhuis}}, \bibinfo {author} {\bibfnamefont {D.}~\bibnamefont {Chandler}},
  \bibinfo {author} {\bibfnamefont {C.}~\bibnamefont {Dellago}}, \ and\
  \bibinfo {author} {\bibfnamefont {P.~L.}\ \bibnamefont {Geissler}},\
  }\href@noop {} {\bibfield  {journal} {\bibinfo  {journal} {Ann. Rev. Phys.
  Chem.}\ }\textbf {\bibinfo {volume} {53}},\ \bibinfo {pages} {291} (\bibinfo
  {year} {2002})}\BibitemShut {NoStop}%
\bibitem [{\citenamefont {ben Avraham}\ and\ \citenamefont
  {Havlin}(2000)}]{benAvraham2000}%
  \BibitemOpen
  \bibfield  {author} {\bibinfo {author} {\bibfnamefont {D.}~\bibnamefont {ben
  Avraham}}\ and\ \bibinfo {author} {\bibfnamefont {S.}~\bibnamefont
  {Havlin}},\ }\href@noop {} {\emph {\bibinfo {title} {Diffusion and Reactions
  in Fractals and Disordered Systems}}}\ (\bibinfo  {publisher} {Cambridge},\
  \bibinfo {address} {Cambridge},\ \bibinfo {year} {2000})\BibitemShut
  {NoStop}%
\bibitem [{\citenamefont {Condamin}\ \emph {et~al.}(2007)\citenamefont
  {Condamin}, \citenamefont {B{\'{e}}nichou}, \citenamefont {Tejedor},
  \citenamefont {Voituriez},\ and\ \citenamefont {Klafter}}]{Condamin2007}%
  \BibitemOpen
  \bibfield  {author} {\bibinfo {author} {\bibfnamefont {S.}~\bibnamefont
  {Condamin}}, \bibinfo {author} {\bibfnamefont {O.}~\bibnamefont
  {B{\'{e}}nichou}}, \bibinfo {author} {\bibfnamefont {V.}~\bibnamefont
  {Tejedor}}, \bibinfo {author} {\bibfnamefont {R.}~\bibnamefont {Voituriez}},
  \ and\ \bibinfo {author} {\bibfnamefont {J.}~\bibnamefont {Klafter}},\
  }\href@noop {} {\bibfield  {journal} {\bibinfo  {journal} {Nature}\ }\textbf
  {\bibinfo {volume} {450}},\ \bibinfo {pages} {77} (\bibinfo {year}
  {2007})}\BibitemShut {NoStop}%
\bibitem [{\citenamefont {Roma}\ \emph {et~al.}(2005)\citenamefont {Roma},
  \citenamefont {O'Flanagan}, \citenamefont {Ruckenstein},\ and\ \citenamefont
  {Sengupta}}]{Roma2005}%
  \BibitemOpen
  \bibfield  {author} {\bibinfo {author} {\bibfnamefont {D.~M.}\ \bibnamefont
  {Roma}}, \bibinfo {author} {\bibfnamefont {R.~A.}\ \bibnamefont
  {O'Flanagan}}, \bibinfo {author} {\bibfnamefont {A.~E.}\ \bibnamefont
  {Ruckenstein}}, \ and\ \bibinfo {author} {\bibfnamefont {A.~M.}\ \bibnamefont
  {Sengupta}},\ }\href@noop {} {\bibfield  {journal} {\bibinfo  {journal}
  {Phys. Rev. E}\ }\textbf {\bibinfo {volume} {71}},\ \bibinfo {pages} {011902}
  (\bibinfo {year} {2005})}\BibitemShut {NoStop}%
\bibitem [{\citenamefont {Waddington}(1957)}]{Waddington1957}%
  \BibitemOpen
  \bibfield  {author} {\bibinfo {author} {\bibfnamefont {C.~H.}\ \bibnamefont
  {Waddington}},\ }\href@noop {} {\emph {\bibinfo {title} {The Strategy of the
  Genes. A Discussion of Some Aspects of Theoretical Biology}}}\ (\bibinfo
  {publisher} {Allen and Unwin},\ \bibinfo {address} {London},\ \bibinfo {year}
  {1957})\BibitemShut {NoStop}%
\bibitem [{\citenamefont {Kauffman}(1993)}]{Kauffman1993}%
  \BibitemOpen
  \bibfield  {author} {\bibinfo {author} {\bibfnamefont {S.}~\bibnamefont
  {Kauffman}},\ }\href@noop {} {\emph {\bibinfo {title} {The Origins of Order:
  Self-Organization and Selection in Evolution}}}\ (\bibinfo  {publisher}
  {Oxford University Press},\ \bibinfo {address} {New York},\ \bibinfo {year}
  {1993})\BibitemShut {NoStop}%
\bibitem [{\citenamefont {Enver}\ \emph {et~al.}(2009)\citenamefont {Enver},
  \citenamefont {Pera}, \citenamefont {Peterson},\ and\ \citenamefont
  {Andrews}}]{Enver2009}%
  \BibitemOpen
  \bibfield  {author} {\bibinfo {author} {\bibfnamefont {T.}~\bibnamefont
  {Enver}}, \bibinfo {author} {\bibfnamefont {M.}~\bibnamefont {Pera}},
  \bibinfo {author} {\bibfnamefont {C.}~\bibnamefont {Peterson}}, \ and\
  \bibinfo {author} {\bibfnamefont {P.~W.}\ \bibnamefont {Andrews}},\
  }\href@noop {} {\bibfield  {journal} {\bibinfo  {journal} {Cell Stem Cell}\
  }\textbf {\bibinfo {volume} {4}},\ \bibinfo {pages} {387} (\bibinfo {year}
  {2009})}\BibitemShut {NoStop}%
\bibitem [{\citenamefont {Bridgham}\ \emph {et~al.}(2009)\citenamefont
  {Bridgham}, \citenamefont {Ortlund},\ and\ \citenamefont
  {Thornton}}]{Bridgham2009}%
  \BibitemOpen
  \bibfield  {author} {\bibinfo {author} {\bibfnamefont {J.~T.}\ \bibnamefont
  {Bridgham}}, \bibinfo {author} {\bibfnamefont {E.~A.}\ \bibnamefont
  {Ortlund}}, \ and\ \bibinfo {author} {\bibfnamefont {J.~W.}\ \bibnamefont
  {Thornton}},\ }\href@noop {} {\bibfield  {journal} {\bibinfo  {journal}
  {Nature}\ }\textbf {\bibinfo {volume} {461}},\ \bibinfo {pages} {515}
  (\bibinfo {year} {2009})}\BibitemShut {NoStop}%
\bibitem [{\citenamefont {Lobkovsky}\ \emph {et~al.}(2011)\citenamefont
  {Lobkovsky}, \citenamefont {Wolf},\ and\ \citenamefont
  {Koonin}}]{Lobkovsky2011}%
  \BibitemOpen
  \bibfield  {author} {\bibinfo {author} {\bibfnamefont {A.~E.}\ \bibnamefont
  {Lobkovsky}}, \bibinfo {author} {\bibfnamefont {Y.~I.}\ \bibnamefont {Wolf}},
  \ and\ \bibinfo {author} {\bibfnamefont {E.~V.}\ \bibnamefont {Koonin}},\
  }\href@noop {} {\bibfield  {journal} {\bibinfo  {journal} {PLoS Comput.
  Biol.}\ }\textbf {\bibinfo {volume} {7}},\ \bibinfo {pages} {e1002302}
  (\bibinfo {year} {2011})}\BibitemShut {NoStop}%
\bibitem [{\citenamefont {Noh}\ and\ \citenamefont {Rieger}(2004)}]{Noh2004}%
  \BibitemOpen
  \bibfield  {author} {\bibinfo {author} {\bibfnamefont {J.~D.}\ \bibnamefont
  {Noh}}\ and\ \bibinfo {author} {\bibfnamefont {H.}~\bibnamefont {Rieger}},\
  }\href@noop {} {\bibfield  {journal} {\bibinfo  {journal} {Phys. Rev. Lett.}\
  }\textbf {\bibinfo {volume} {92}},\ \bibinfo {pages} {118701} (\bibinfo
  {year} {2004})}\BibitemShut {NoStop}%
\bibitem [{\citenamefont {Bollt}\ and\ \citenamefont {ben
  Avraham}(2005)}]{Bollt2005}%
  \BibitemOpen
  \bibfield  {author} {\bibinfo {author} {\bibfnamefont {E.~M.}\ \bibnamefont
  {Bollt}}\ and\ \bibinfo {author} {\bibfnamefont {D.}~\bibnamefont {ben
  Avraham}},\ }\href@noop {} {\bibfield  {journal} {\bibinfo  {journal} {N. J.
  Phys.}\ }\textbf {\bibinfo {volume} {7}},\ \bibinfo {pages} {26} (\bibinfo
  {year} {2005})}\BibitemShut {NoStop}%
\bibitem [{\citenamefont {Dellago}\ \emph {et~al.}(1998)\citenamefont
  {Dellago}, \citenamefont {Bolhuis}, \citenamefont {Csajka},\ and\
  \citenamefont {Chandler}}]{Dellago1998}%
  \BibitemOpen
  \bibfield  {author} {\bibinfo {author} {\bibfnamefont {C.}~\bibnamefont
  {Dellago}}, \bibinfo {author} {\bibfnamefont {P.~G.}\ \bibnamefont
  {Bolhuis}}, \bibinfo {author} {\bibfnamefont {F.~S.}\ \bibnamefont {Csajka}},
  \ and\ \bibinfo {author} {\bibfnamefont {D.}~\bibnamefont {Chandler}},\
  }\href@noop {} {\bibfield  {journal} {\bibinfo  {journal} {J. Chem. Phys.}\
  }\textbf {\bibinfo {volume} {108}},\ \bibinfo {pages} {1964} (\bibinfo {year}
  {1998})}\BibitemShut {NoStop}%
\bibitem [{\citenamefont {Dellago}\ \emph {et~al.}(2003)\citenamefont
  {Dellago}, \citenamefont {Bolhuis},\ and\ \citenamefont
  {Geissler}}]{Dellago2003}%
  \BibitemOpen
  \bibfield  {author} {\bibinfo {author} {\bibfnamefont {C.}~\bibnamefont
  {Dellago}}, \bibinfo {author} {\bibfnamefont {P.~G.}\ \bibnamefont
  {Bolhuis}}, \ and\ \bibinfo {author} {\bibfnamefont {P.~L.}\ \bibnamefont
  {Geissler}},\ }\href@noop {} {\bibfield  {journal} {\bibinfo  {journal} {Adv.
  Chem. Phys.}\ }\textbf {\bibinfo {volume} {123}},\ \bibinfo {pages} {1}
  (\bibinfo {year} {2003})}\BibitemShut {NoStop}%
\bibitem [{\citenamefont {Hummer}(2004)}]{Hummer2004}%
  \BibitemOpen
  \bibfield  {author} {\bibinfo {author} {\bibfnamefont {G.}~\bibnamefont
  {Hummer}},\ }\href@noop {} {\bibfield  {journal} {\bibinfo  {journal} {J.
  Chem. Phys.}\ }\textbf {\bibinfo {volume} {120}},\ \bibinfo {pages} {516}
  (\bibinfo {year} {2004})}\BibitemShut {NoStop}%
\bibitem [{\citenamefont {Harland}\ and\ \citenamefont
  {Sun}(2007)}]{Harland2007}%
  \BibitemOpen
  \bibfield  {author} {\bibinfo {author} {\bibfnamefont {B.}~\bibnamefont
  {Harland}}\ and\ \bibinfo {author} {\bibfnamefont {S.~X.}\ \bibnamefont
  {Sun}},\ }\href@noop {} {\bibfield  {journal} {\bibinfo  {journal} {J. Chem.
  Phys.}\ }\textbf {\bibinfo {volume} {127}},\ \bibinfo {pages} {104103}
  (\bibinfo {year} {2007})}\BibitemShut {NoStop}%
\bibitem [{\citenamefont {E}\ and\ \citenamefont
  {Vanden-Eijnden}(2006)}]{E2006}%
  \BibitemOpen
  \bibfield  {author} {\bibinfo {author} {\bibfnamefont {W.}~\bibnamefont {E}}\
  and\ \bibinfo {author} {\bibfnamefont {E.}~\bibnamefont {Vanden-Eijnden}},\
  }\href@noop {} {\bibfield  {journal} {\bibinfo  {journal} {J. Stat. Phys.}\
  }\textbf {\bibinfo {volume} {123}},\ \bibinfo {pages} {503} (\bibinfo {year}
  {2006})}\BibitemShut {NoStop}%
\bibitem [{\citenamefont {Metzner}\ \emph {et~al.}(2009)\citenamefont
  {Metzner}, \citenamefont {Sch{\"{u}}tte},\ and\ \citenamefont
  {Vanden-Eijnden}}]{Metzner2009}%
  \BibitemOpen
  \bibfield  {author} {\bibinfo {author} {\bibfnamefont {P.}~\bibnamefont
  {Metzner}}, \bibinfo {author} {\bibfnamefont {C.}~\bibnamefont
  {Sch{\"{u}}tte}}, \ and\ \bibinfo {author} {\bibfnamefont {E.}~\bibnamefont
  {Vanden-Eijnden}},\ }\href@noop {} {\bibfield  {journal} {\bibinfo  {journal}
  {Multiscale Model. Simul.}\ }\textbf {\bibinfo {volume} {7}},\ \bibinfo
  {pages} {1192} (\bibinfo {year} {2009})}\BibitemShut {NoStop}%
\bibitem [{\citenamefont {Weiss}(1994)}]{Weiss1994}%
  \BibitemOpen
  \bibfield  {author} {\bibinfo {author} {\bibfnamefont {G.~H.}\ \bibnamefont
  {Weiss}},\ }\href@noop {} {\emph {\bibinfo {title} {Aspects and Applications
  of the Random Walk}}}\ (\bibinfo  {publisher} {North Holland},\ \bibinfo
  {address} {Amsterdam},\ \bibinfo {year} {1994})\BibitemShut {NoStop}%
\bibitem [{\citenamefont {Zeldovich}\ \emph {et~al.}(2007)\citenamefont
  {Zeldovich}, \citenamefont {Chen},\ and\ \citenamefont
  {Shakhnovich}}]{Zeldovich2007}%
  \BibitemOpen
  \bibfield  {author} {\bibinfo {author} {\bibfnamefont {K.~B.}\ \bibnamefont
  {Zeldovich}}, \bibinfo {author} {\bibfnamefont {P.}~\bibnamefont {Chen}}, \
  and\ \bibinfo {author} {\bibfnamefont {E.~I.}\ \bibnamefont {Shakhnovich}},\
  }\href@noop {} {\bibfield  {journal} {\bibinfo  {journal} {Proc. Natl. Acad.
  Sci. USA}\ }\textbf {\bibinfo {volume} {104}},\ \bibinfo {pages} {16152}
  (\bibinfo {year} {2007})}\BibitemShut {NoStop}%
\bibitem [{\citenamefont {Tokuriki}\ \emph {et~al.}(2008)\citenamefont
  {Tokuriki}, \citenamefont {Stricher}, \citenamefont {Serrano},\ and\
  \citenamefont {Tawfik}}]{Tokuriki2008}%
  \BibitemOpen
  \bibfield  {author} {\bibinfo {author} {\bibfnamefont {N.}~\bibnamefont
  {Tokuriki}}, \bibinfo {author} {\bibfnamefont {F.}~\bibnamefont {Stricher}},
  \bibinfo {author} {\bibfnamefont {L.}~\bibnamefont {Serrano}}, \ and\
  \bibinfo {author} {\bibfnamefont {D.~S.}\ \bibnamefont {Tawfik}},\
  }\href@noop {} {\bibfield  {journal} {\bibinfo  {journal} {PLoS Comput.
  Biol.}\ }\textbf {\bibinfo {volume} {4}},\ \bibinfo {pages} {e1000002}
  (\bibinfo {year} {2008})}\BibitemShut {NoStop}%
\bibitem [{\citenamefont {Tokuriki}\ and\ \citenamefont
  {Tawfik}(2009)}]{Tokuriki2009}%
  \BibitemOpen
  \bibfield  {author} {\bibinfo {author} {\bibfnamefont {N.}~\bibnamefont
  {Tokuriki}}\ and\ \bibinfo {author} {\bibfnamefont {D.~S.}\ \bibnamefont
  {Tawfik}},\ }\href@noop {} {\bibfield  {journal} {\bibinfo  {journal} {Curr.
  Opin. Struct. Biol.}\ }\textbf {\bibinfo {volume} {19}},\ \bibinfo {pages}
  {596} (\bibinfo {year} {2009})}\BibitemShut {NoStop}%
\bibitem [{\citenamefont {Bloom}\ and\ \citenamefont
  {Arnold}(2009)}]{Bloom2009}%
  \BibitemOpen
  \bibfield  {author} {\bibinfo {author} {\bibfnamefont {J.~D.}\ \bibnamefont
  {Bloom}}\ and\ \bibinfo {author} {\bibfnamefont {F.~H.}\ \bibnamefont
  {Arnold}},\ }\href@noop {} {\bibfield  {journal} {\bibinfo  {journal} {Proc.
  Natl. Acad. Sci. USA}\ }\textbf {\bibinfo {volume} {106}},\ \bibinfo {pages}
  {9995} (\bibinfo {year} {2009})}\BibitemShut {NoStop}%
\bibitem [{\citenamefont {Whitehead}\ \emph {et~al.}(2012)\citenamefont
  {Whitehead}, \citenamefont {Chevalier}, \citenamefont {Song}, \citenamefont
  {Dreyfus}, \citenamefont {Fleishman}, \citenamefont {{De Mattos}},
  \citenamefont {Myers}, \citenamefont {Kamisetty}, \citenamefont {Blair},
  \citenamefont {Wilson},\ and\ \citenamefont {Baker}}]{Whitehead2012}%
  \BibitemOpen
  \bibfield  {author} {\bibinfo {author} {\bibfnamefont {T.~A.}\ \bibnamefont
  {Whitehead}}, \bibinfo {author} {\bibfnamefont {A.}~\bibnamefont
  {Chevalier}}, \bibinfo {author} {\bibfnamefont {Y.}~\bibnamefont {Song}},
  \bibinfo {author} {\bibfnamefont {C.}~\bibnamefont {Dreyfus}}, \bibinfo
  {author} {\bibfnamefont {S.~J.}\ \bibnamefont {Fleishman}}, \bibinfo {author}
  {\bibfnamefont {C.}~\bibnamefont {{De Mattos}}}, \bibinfo {author}
  {\bibfnamefont {C.~A.}\ \bibnamefont {Myers}}, \bibinfo {author}
  {\bibfnamefont {H.}~\bibnamefont {Kamisetty}}, \bibinfo {author}
  {\bibfnamefont {P.}~\bibnamefont {Blair}}, \bibinfo {author} {\bibfnamefont
  {I.~A.}\ \bibnamefont {Wilson}}, \ and\ \bibinfo {author} {\bibfnamefont
  {D.}~\bibnamefont {Baker}},\ }\href@noop {} {\bibfield  {journal} {\bibinfo
  {journal} {Nature Biotech.}\ }\textbf {\bibinfo {volume} {30}},\ \bibinfo
  {pages} {543} (\bibinfo {year} {2012})}\BibitemShut {NoStop}%
\bibitem [{\citenamefont {H{\"{a}}nggi}\ \emph {et~al.}(1990)\citenamefont
  {H{\"{a}}nggi}, \citenamefont {Talkner},\ and\ \citenamefont
  {Borkovec}}]{Hanggi1990}%
  \BibitemOpen
  \bibfield  {author} {\bibinfo {author} {\bibfnamefont {P.}~\bibnamefont
  {H{\"{a}}nggi}}, \bibinfo {author} {\bibfnamefont {P.}~\bibnamefont
  {Talkner}}, \ and\ \bibinfo {author} {\bibfnamefont {M.}~\bibnamefont
  {Borkovec}},\ }\href@noop {} {\bibfield  {journal} {\bibinfo  {journal} {Rev.
  Mod. Phys.}\ }\textbf {\bibinfo {volume} {62}},\ \bibinfo {pages} {251}
  (\bibinfo {year} {1990})}\BibitemShut {NoStop}%
\bibitem [{\citenamefont {Metzner}\ \emph {et~al.}(2006)\citenamefont
  {Metzner}, \citenamefont {Sch{\"{u}}tte},\ and\ \citenamefont
  {Vanden-Eijnden}}]{Metzner2006}%
  \BibitemOpen
  \bibfield  {author} {\bibinfo {author} {\bibfnamefont {P.}~\bibnamefont
  {Metzner}}, \bibinfo {author} {\bibfnamefont {C.}~\bibnamefont
  {Sch{\"{u}}tte}}, \ and\ \bibinfo {author} {\bibfnamefont {E.}~\bibnamefont
  {Vanden-Eijnden}},\ }\href@noop {} {\bibfield  {journal} {\bibinfo  {journal}
  {J. Chem. Phys.}\ }\textbf {\bibinfo {volume} {125}},\ \bibinfo {pages}
  {084110} (\bibinfo {year} {2006})}\BibitemShut {NoStop}%
\bibitem [{\citenamefont {Mayer}\ \emph {et~al.}(2007)\citenamefont {Mayer},
  \citenamefont {R{\"{u}}diger}, \citenamefont {Ang}, \citenamefont {Joerger},\
  and\ \citenamefont {Fersht}}]{Mayer2007}%
  \BibitemOpen
  \bibfield  {author} {\bibinfo {author} {\bibfnamefont {S.}~\bibnamefont
  {Mayer}}, \bibinfo {author} {\bibfnamefont {S.}~\bibnamefont
  {R{\"{u}}diger}}, \bibinfo {author} {\bibfnamefont {H.~C.}\ \bibnamefont
  {Ang}}, \bibinfo {author} {\bibfnamefont {A.~C.}\ \bibnamefont {Joerger}}, \
  and\ \bibinfo {author} {\bibfnamefont {A.~R.}\ \bibnamefont {Fersht}},\
  }\href@noop {} {\bibfield  {journal} {\bibinfo  {journal} {J. Mol. Biol.}\
  }\textbf {\bibinfo {volume} {372}},\ \bibinfo {pages} {268} (\bibinfo {year}
  {2007})}\BibitemShut {NoStop}%
\bibitem [{\citenamefont {Clackson}\ and\ \citenamefont
  {Wells}(1995)}]{Clackson1995}%
  \BibitemOpen
  \bibfield  {author} {\bibinfo {author} {\bibfnamefont {T.}~\bibnamefont
  {Clackson}}\ and\ \bibinfo {author} {\bibfnamefont {J.~A.}\ \bibnamefont
  {Wells}},\ }\href@noop {} {\bibfield  {journal} {\bibinfo  {journal}
  {Science}\ }\textbf {\bibinfo {volume} {267}},\ \bibinfo {pages} {383}
  (\bibinfo {year} {1995})}\BibitemShut {NoStop}%
\bibitem [{\citenamefont {Serrano}\ \emph {et~al.}(1993)\citenamefont
  {Serrano}, \citenamefont {Day},\ and\ \citenamefont {Fersht}}]{Serrano1993}%
  \BibitemOpen
  \bibfield  {author} {\bibinfo {author} {\bibfnamefont {L.}~\bibnamefont
  {Serrano}}, \bibinfo {author} {\bibfnamefont {A.~G.}\ \bibnamefont {Day}}, \
  and\ \bibinfo {author} {\bibfnamefont {A.~R.}\ \bibnamefont {Fersht}},\
  }\href@noop {} {\bibfield  {journal} {\bibinfo  {journal} {J. Mol. Biol.}\
  }\textbf {\bibinfo {volume} {233}},\ \bibinfo {pages} {305} (\bibinfo {year}
  {1993})}\BibitemShut {NoStop}%
\bibitem [{\citenamefont {Tokuriki}\ \emph {et~al.}(2007)\citenamefont
  {Tokuriki}, \citenamefont {Stricher}, \citenamefont {Schymkowitz},
  \citenamefont {Serrano},\ and\ \citenamefont {Tawfik}}]{Tokuriki2007}%
  \BibitemOpen
  \bibfield  {author} {\bibinfo {author} {\bibfnamefont {N.}~\bibnamefont
  {Tokuriki}}, \bibinfo {author} {\bibfnamefont {F.}~\bibnamefont {Stricher}},
  \bibinfo {author} {\bibfnamefont {J.}~\bibnamefont {Schymkowitz}}, \bibinfo
  {author} {\bibfnamefont {L.}~\bibnamefont {Serrano}}, \ and\ \bibinfo
  {author} {\bibfnamefont {D.~S.}\ \bibnamefont {Tawfik}},\ }\href@noop {}
  {\bibfield  {journal} {\bibinfo  {journal} {J. Mol. Biol.}\ }\textbf
  {\bibinfo {volume} {369}},\ \bibinfo {pages} {1318} (\bibinfo {year}
  {2007})}\BibitemShut {NoStop}%
\bibitem [{\citenamefont {Bogan}\ and\ \citenamefont
  {Thorn}(1998)}]{Bogan1998}%
  \BibitemOpen
  \bibfield  {author} {\bibinfo {author} {\bibfnamefont {A.~A.}\ \bibnamefont
  {Bogan}}\ and\ \bibinfo {author} {\bibfnamefont {K.~S.}\ \bibnamefont
  {Thorn}},\ }\href@noop {} {\bibfield  {journal} {\bibinfo  {journal} {J. Mol.
  Biol.}\ }\textbf {\bibinfo {volume} {280}},\ \bibinfo {pages} {1} (\bibinfo
  {year} {1998})}\BibitemShut {NoStop}%
\bibitem [{\citenamefont {Thorn}\ and\ \citenamefont
  {Bogan}(2001)}]{Thorn2001}%
  \BibitemOpen
  \bibfield  {author} {\bibinfo {author} {\bibfnamefont {K.~S.}\ \bibnamefont
  {Thorn}}\ and\ \bibinfo {author} {\bibfnamefont {A.~A.}\ \bibnamefont
  {Bogan}},\ }\href@noop {} {\bibfield  {journal} {\bibinfo  {journal}
  {Bioinformatics}\ }\textbf {\bibinfo {volume} {17}},\ \bibinfo {pages} {284}
  (\bibinfo {year} {2001})}\BibitemShut {NoStop}%
\bibitem [{\citenamefont {Wang}\ \emph {et~al.}(2002)\citenamefont {Wang},
  \citenamefont {Minasov},\ and\ \citenamefont {Shoichet}}]{Wang2002}%
  \BibitemOpen
  \bibfield  {author} {\bibinfo {author} {\bibfnamefont {X.}~\bibnamefont
  {Wang}}, \bibinfo {author} {\bibfnamefont {G.}~\bibnamefont {Minasov}}, \
  and\ \bibinfo {author} {\bibfnamefont {B.~K.}\ \bibnamefont {Shoichet}},\
  }\href@noop {} {\bibfield  {journal} {\bibinfo  {journal} {J. Mol. Biol.}\
  }\textbf {\bibinfo {volume} {320}},\ \bibinfo {pages} {85} (\bibinfo {year}
  {2002})}\BibitemShut {NoStop}%
\bibitem [{\citenamefont {Crow}\ and\ \citenamefont {Kimura}(1970)}]{Crow1970}%
  \BibitemOpen
  \bibfield  {author} {\bibinfo {author} {\bibfnamefont {J.~F.}\ \bibnamefont
  {Crow}}\ and\ \bibinfo {author} {\bibfnamefont {M.}~\bibnamefont {Kimura}},\
  }\href@noop {} {\emph {\bibinfo {title} {An Introduction to Population
  Genetics Theory}}}\ (\bibinfo  {publisher} {Harper and Row},\ \bibinfo
  {address} {New York},\ \bibinfo {year} {1970})\BibitemShut {NoStop}%
\bibitem [{\citenamefont {Manhart}\ \emph {et~al.}(2012)\citenamefont
  {Manhart}, \citenamefont {Haldane},\ and\ \citenamefont
  {Morozov}}]{Manhart2012}%
  \BibitemOpen
  \bibfield  {author} {\bibinfo {author} {\bibfnamefont {M.}~\bibnamefont
  {Manhart}}, \bibinfo {author} {\bibfnamefont {A.}~\bibnamefont {Haldane}}, \
  and\ \bibinfo {author} {\bibfnamefont {A.~V.}\ \bibnamefont {Morozov}},\
  }\href@noop {} {\bibfield  {journal} {\bibinfo  {journal} {Theor. Pop.
  Biol.}\ }\textbf {\bibinfo {volume} {82}},\ \bibinfo {pages} {66} (\bibinfo
  {year} {2012})}\BibitemShut {NoStop}%
\bibitem [{\citenamefont {DePristo}\ \emph {et~al.}(2005)\citenamefont
  {DePristo}, \citenamefont {Weinreich},\ and\ \citenamefont
  {Hartl}}]{DePristo2005}%
  \BibitemOpen
  \bibfield  {author} {\bibinfo {author} {\bibfnamefont {M.~A.}\ \bibnamefont
  {DePristo}}, \bibinfo {author} {\bibfnamefont {D.~M.}\ \bibnamefont
  {Weinreich}}, \ and\ \bibinfo {author} {\bibfnamefont {D.~L.}\ \bibnamefont
  {Hartl}},\ }\href@noop {} {\bibfield  {journal} {\bibinfo  {journal} {Nat.
  Rev. Genet.}\ }\textbf {\bibinfo {volume} {6}},\ \bibinfo {pages} {678}
  (\bibinfo {year} {2005})}\BibitemShut {NoStop}%
\bibitem [{\citenamefont {Bershtein}\ \emph {et~al.}(2008)\citenamefont
  {Bershtein}, \citenamefont {Goldin},\ and\ \citenamefont
  {Tawfik}}]{Bershtein2008}%
  \BibitemOpen
  \bibfield  {author} {\bibinfo {author} {\bibfnamefont {S.}~\bibnamefont
  {Bershtein}}, \bibinfo {author} {\bibfnamefont {K.}~\bibnamefont {Goldin}}, \
  and\ \bibinfo {author} {\bibfnamefont {D.~S.}\ \bibnamefont {Tawfik}},\
  }\href@noop {} {\bibfield  {journal} {\bibinfo  {journal} {J. Mol. Biol.}\
  }\textbf {\bibinfo {volume} {379}},\ \bibinfo {pages} {1029} (\bibinfo {year}
  {2008})}\BibitemShut {NoStop}%
\bibitem [{\citenamefont {Brown}\ \emph {et~al.}(2011)\citenamefont {Brown},
  \citenamefont {Johnson}, \citenamefont {Dunker},\ and\ \citenamefont
  {Daughdrill}}]{Brown2011}%
  \BibitemOpen
  \bibfield  {author} {\bibinfo {author} {\bibfnamefont {C.~J.}\ \bibnamefont
  {Brown}}, \bibinfo {author} {\bibfnamefont {A.~K.}\ \bibnamefont {Johnson}},
  \bibinfo {author} {\bibfnamefont {A.~K.}\ \bibnamefont {Dunker}}, \ and\
  \bibinfo {author} {\bibfnamefont {G.~W.}\ \bibnamefont {Daughdrill}},\
  }\href@noop {} {\bibfield  {journal} {\bibinfo  {journal} {Curr. Opin.
  Struct. Biol.}\ }\textbf {\bibinfo {volume} {21}},\ \bibinfo {pages} {441}
  (\bibinfo {year} {2011})}\BibitemShut {NoStop}%
\bibitem [{\citenamefont {Rutherford}(2003)}]{Rutherford2003}%
  \BibitemOpen
  \bibfield  {author} {\bibinfo {author} {\bibfnamefont {S.~L.}\ \bibnamefont
  {Rutherford}},\ }\href@noop {} {\bibfield  {journal} {\bibinfo  {journal}
  {Nat. Rev. Genet.}\ }\textbf {\bibinfo {volume} {4}},\ \bibinfo {pages} {263}
  (\bibinfo {year} {2003})}\BibitemShut {NoStop}%
\end{thebibliography}%

\newpage

\onecolumngrid

\begin{center}
\LARGE{Supplementary Material: \\ A path-based approach to random walks on networks \\ characterizes how proteins evolve new function}

\vspace{0.5cm}

\large{Michael Manhart$^1$ and Alexandre V. Morozov$^{1,2}$}

\vspace{0.5cm}

	\small{\emph{$^1$ Department of Physics and Astronomy, Rutgers University, Piscataway, NJ 08854}}
	
	\small{\emph{$^2$ BioMaPS Institute for Quantitative Biology, Rutgers University, Piscataway, NJ 08854}}
\end{center}

\renewcommand{\thefigure}{S\arabic{figure}}
\setcounter{figure}{0}

\normalsize

\begin{figure}[h]
\begin{center}
\includegraphics[scale=0.345]{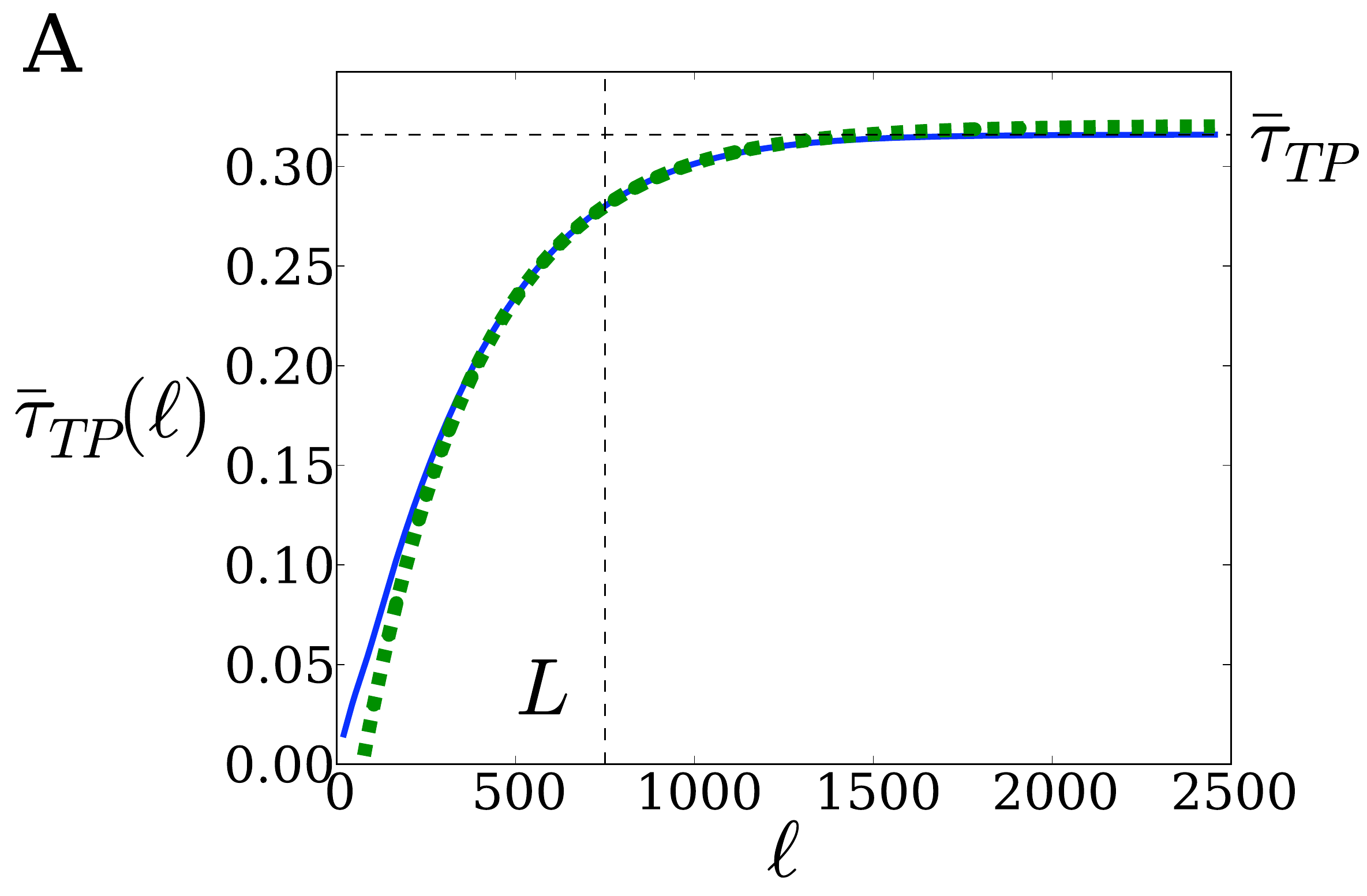}
\includegraphics[scale=0.34]{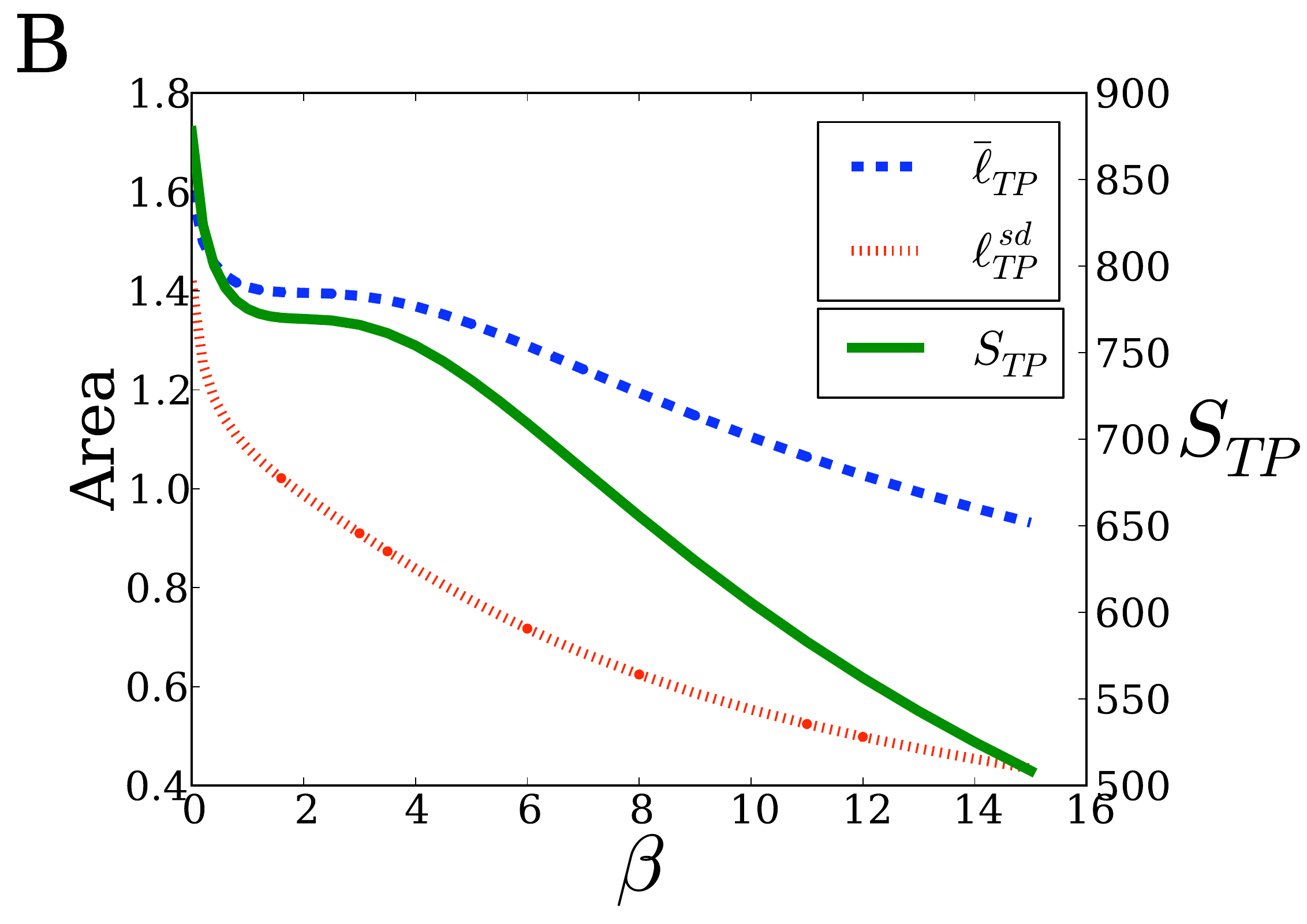}
\end{center}
\caption{(A)~For the ensemble of transition paths (TPs) in the 2D double-well potential, the mean time $\bar{\tau}_\TP(\ell)$ of paths up to length $\ell$. In the limit $\ell \arr \infty$, $\bar{\tau}_\TP(\ell)$ converges to the total mean time $\bar{\tau}_\TP$. Similar to $\rho_\TP(\ell)$ in Fig.~1C, for sufficiently large $\ell$ $\bar{\tau}_\TP(\ell)$ acquires a universal exponential form: $(\bar{\tau}_\TP - \bar{\tau}_\TP(\ell)) \sim e^{-a\ell/\bar{\ell}_\TP}$.  We choose $L=750$ as the effective cutoff used to fit an exponential tail of $\bar{\tau}_\TP(\ell)$; fit in the range $\ell \in [L-50, L]$ (dashed, green) closely matches the full calculation (solid, blue) for $\ell > L$.
(B)~The mean length $\bar{\ell}_\TP$ (dashed, blue) and standard deviation $\ell_\TP^\text{sd}$ (dotted, red) as functions of inverse temperature $\beta$. Both quantities have units of area since they are rescaled by $(\Delta x)^2$. Also shown is the entropy $S_\TP$ (solid, green) of TPs.
}
\end{figure}

\begin{figure}
\begin{center}
\includegraphics[scale=1.1]{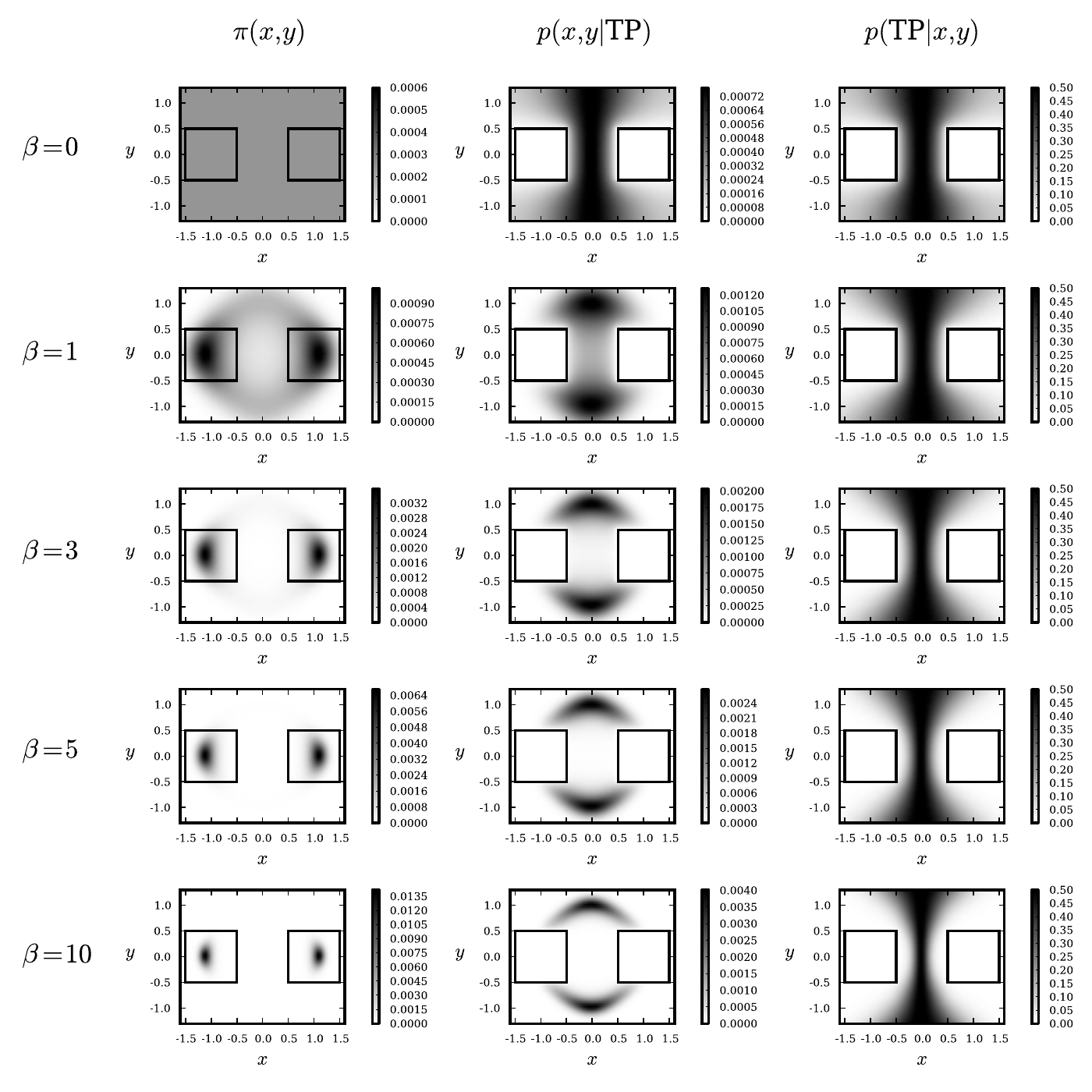}
\end{center}
\caption{For the 2D double-well potential, the equilibrium distribution of states $\pi(x,y)$, density of states on TPs $p(x,y|\TP)$, and TP densities (given that the system is at $(x,y)$, the probability it is on a TP: $p(\TP|x,y) = \average{\mathcal{I}_{(x,y)}}_\TP \Z_\TP/(\Z_\TP \average{\mathcal{I}_{(x,y)}}_\TP + \Z_\RP \average{\mathcal{I}_{(x,y)}}_\RP)$) at different values of $\beta$.  All calculations use $\Delta x = 0.05$.}
\label{fig:densities_vs_beta}
\end{figure}


\begin{figure}
\begin{center}
\includegraphics[scale=0.2]{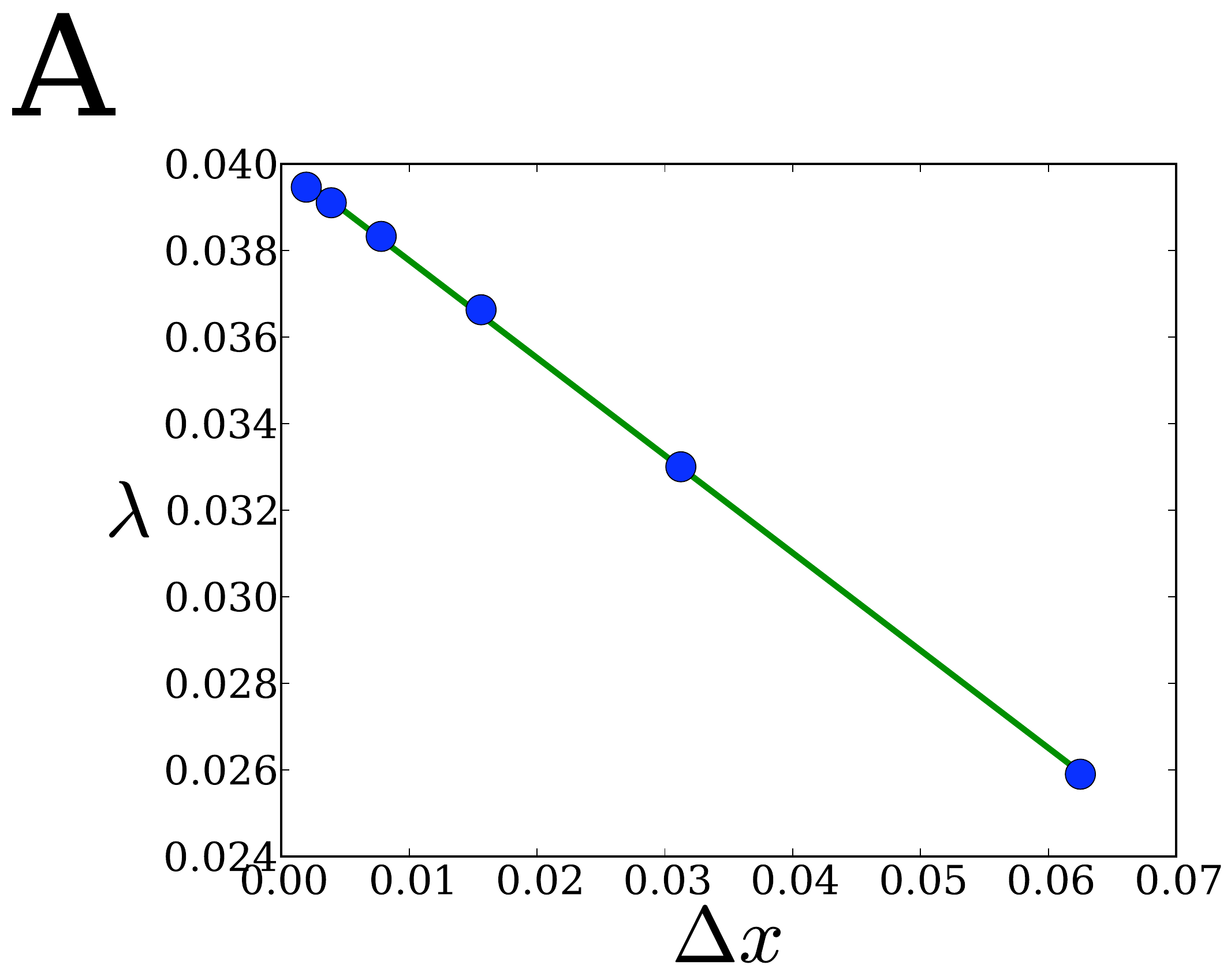} \\
\includegraphics[scale=1]{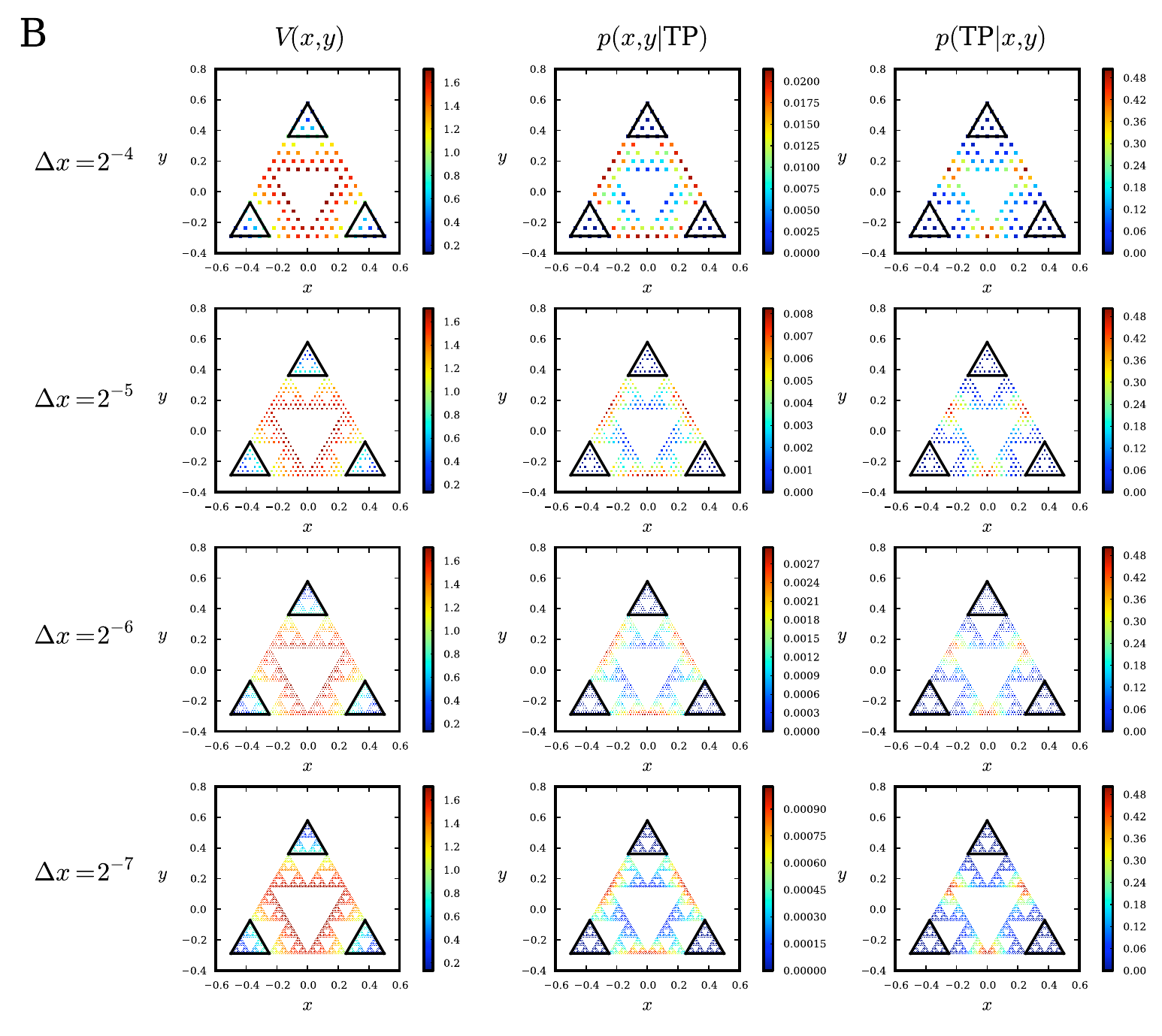}
\end{center}
\caption{Sierpinski triangle embedded in the triple-well potential $V(x,y) = 10\sum_{i=1}^3 ((x-x_i)^2 + (y-y_i)^2)e^{-5(x-x_i)^2 -5(y-y_i)^2}$, where $(x_1,y_1) = (0, 1/\sqrt{3})$, $(x_2,y_2) = (1/2, -1/(2\sqrt{3}))$, and $(x_3,y_3) = (-1/2, -1/(2\sqrt{3}))$.  There are three metastable states outlined in black, one at each corner of the triangle. Monte Carlo jump rates are rescaled by $(\Delta x)^{d_w}$, where $\Delta x = 2^{-n}$ ($n$ is the fractal order) and $d_w = \log 5/\log 2$ is the dimension of a random walk on the Sierpinski triangle. (A) Transition path flux $\lambda$ as a function of lattice spacing $\Delta x$.  As with the double-well potential, analytic continuation of $\lambda(\Delta x)$ allows us to infer the reaction rate $k \approx \lambda/2 \approx 2.0 \times 10^{-2}$ between any pair of metastable states in an infinite-order fractal from a few finite realizations.  (B)~The potential $V(x,y)$, the density of states on TPs $p(x,y|\TP)$, and TP densities $p(\TP|x,y)$ for $\beta = 6$.
}
\label{fig:sierpinski_densities}
\end{figure}


\begin{figure}
\begin{center}
\includegraphics[scale=0.4]{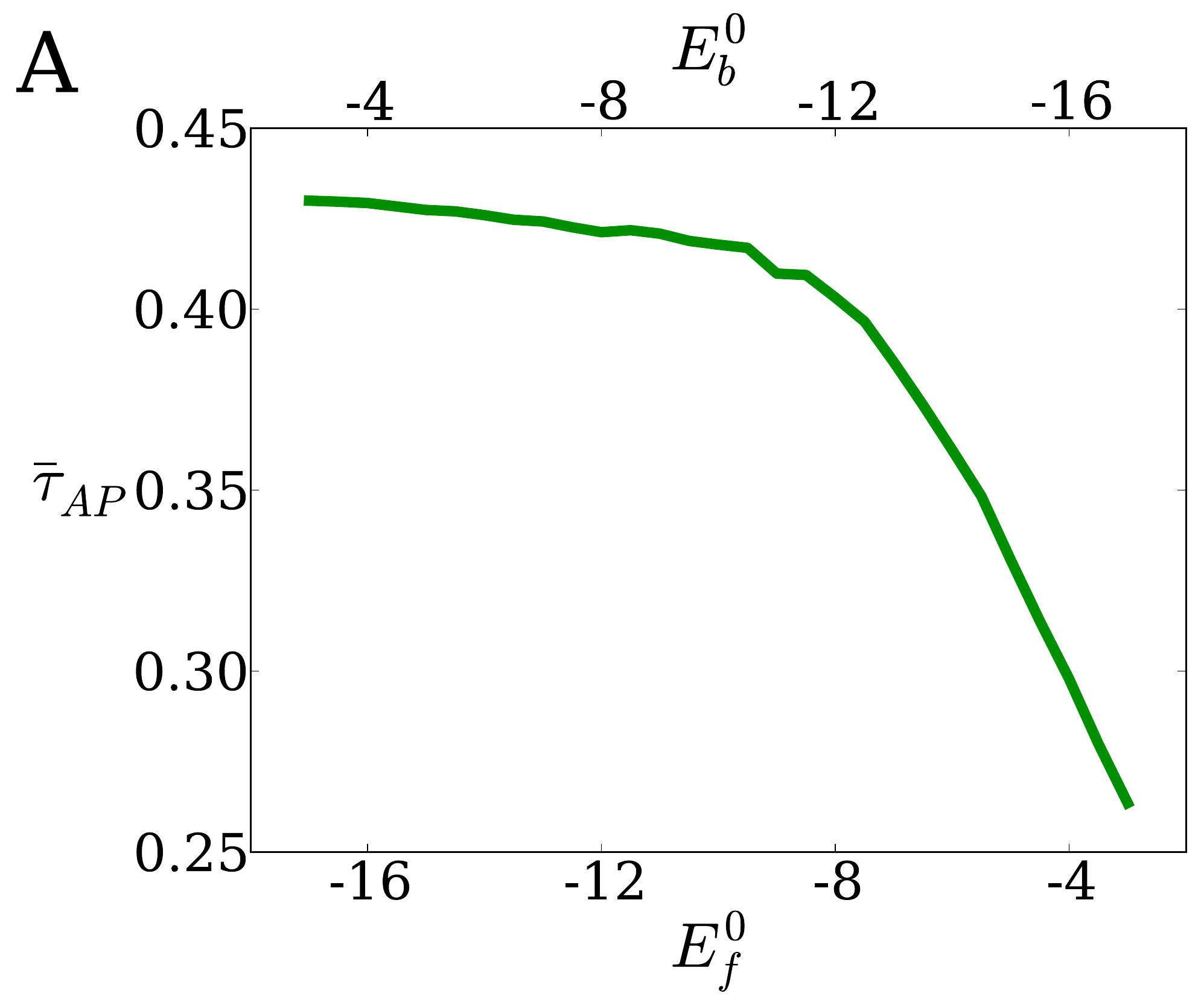}
\includegraphics[scale=0.4]{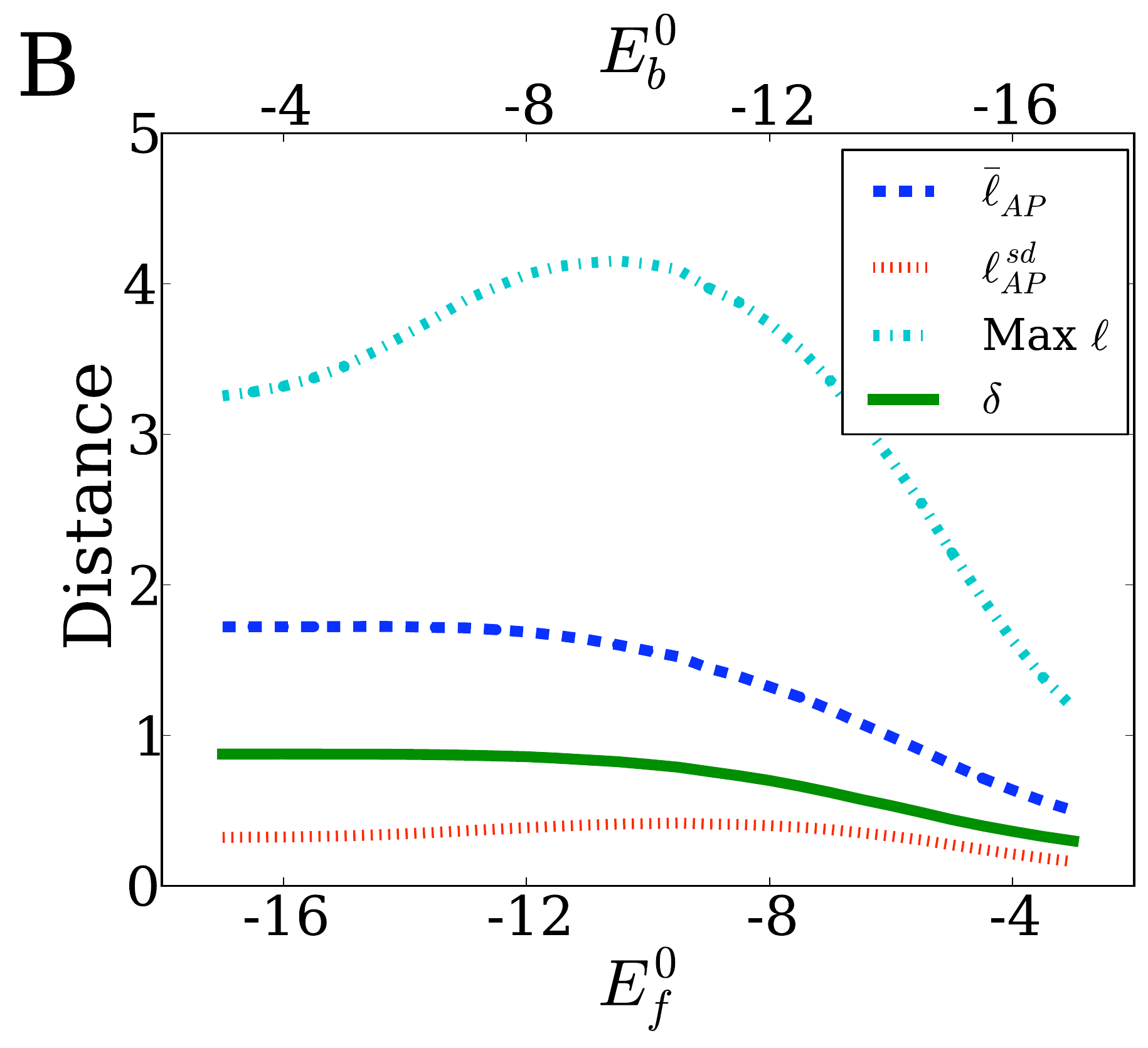}
\end{center}
\caption{(A) Mean time $\bar{\tau}_\AP$ (in units of $(Nu)^{-1}$) of APs for the subspace $E^0_f + E^0_b = -20$ kcal/mol.  (B) Plot of average path length $\bar{\ell}_\AP$ (dashed, blue), standard deviation of path length $\ell_\AP^\text{sd}$ (dotted, red), maximum possible path length (dashed and dotted, cyan), and the average net distance $\delta$ between the initial state and final state (solid, green) for the parameter subspace $E_f^0 + E_b^0 = -20$ kcal/mol.
On average, proteins will undergo twice as many substitutions as the average distance $\delta$.  The maximum number of substitutions is at least $3 \delta$.
The crossover regime allows for the greatest variation in path lengths, both in terms of standard deviation and maximum possible length.  All quantities are per-residue, and data points are averages from $5 \times 10^3$ landscape realizations.}
\label{}
\end{figure}

\end{document}